\documentclass[a4paper,11pt]{article}
\usepackage[pdftex]{graphicx}
\usepackage[T1]{fontenc}
\usepackage{lmodern}
\usepackage{slashed}
\usepackage[utf8]{inputenc}
\usepackage[english]{babel}
\usepackage{microtype}
\usepackage{cite}
\usepackage{amsmath}
\usepackage{amssymb}
\usepackage{amsfonts}
\usepackage{mathtools}
\numberwithin{equation}{section}
\allowdisplaybreaks
\usepackage{geometry}
\usepackage[toc,page]{appendix}
\usepackage[normalem]{ulem}
\usepackage{bm}
\usepackage{appendix}
\usepackage{bbold}
\usepackage{cancel}

\DeclareMathAlphabet{\mathpzc}{OT1}{pzc}{m}{it}
\newcommand{\virgolette}{``}

\begin{document}

\begin{titlepage}
\begin{flushright}
\par\end{flushright}
\vskip 0.5cm
\begin{center}
\textbf{\LARGE \bf Novel Free Differential Algebras for Supergravity}

\vskip 1cm 
\large {\bf Pietro~Antonio~Grassi}$^{~a, b,}$\footnote{pietro.grassi@uniupo.it}\,. 

\vskip .5cm 
{\small
{$^{(a)}$ \it DiSIT,} 
{\it Universit\`a del Piemonte Orientale,} 
{\it viale T.~Michel, 11, 15121 Alessandria, Italy}
\\
{$^{(b)}$ \it INFN, Sezione di Torino}
{\it via P.~Giuria 1, 10125 Torino, Italy} \\
}
\end{center}

\vskip 2cm 
\begin{abstract} 
 We develop the theory of Free Integro-Differential Algebras (FIDA) extending the powerful technique 
 of Free Differential Algebras constructed by D. Sullivan. We extend the analysis beyond the superforms  
 to integral- and pseudo-forms used in supergeometry. It is shown that there are novel structures that might open the road to a deeper understanding of the geometry of supergravity.  
 We apply the technique to some models
 as an illustration and we provide a complete analysis for D=11 supergravity. There, it is shown how the Hodge star operator for supermanifolds can be used to analyze the set of cocycles and to build the corresponding FIDA. A new integral form emerges which plays the role of the truly dual to 4-form $F^{(4)}$ and we propose a 
 new variational principle on supermanifolds. 
 
\end{abstract}

\vfill{}
\vspace{1.5cm}
\end{titlepage}

\setcounter{footnote}{0}
\tableofcontents

\section{Ingredients}

The existence of a single integral form beyond the Free Differential Algebra (FDA) 
is proved and this is the true essence of the 
supergeometric nature of supergravity and our analysis is a confirmation of this fact from a different point of view. The extension of the FDA is named Free Integro-Differential Algebra (FIDA) in the paper. 
 
 As is pointed out by different authors \cite{Wess:1977fn,gm11,gm12,gm13,VanNieuwenhuizen:1981ae,gm14,GGRS,WB,book,Kuzenko:1998tsq} 
 the superspace nature of supersymmetric and supergravity models is intimately related to the supergeometric structure due to vielbeins and gravitinos. Their supersymmetry and diffeomorphism transformations can be recast in a beautiful geometric framework known as {the \it rheonomic} approach. This allows us to 
use the powerful technique of Cartan calculus, such as the exterior differential, the contraction operator, the Lie derivates, etc...
and to compute some cohomologies for flat or curved supermanifolds. It turns out that the Chevalley-Eilenberg 
cohomology is usually non-trivial and it can be conveniently understood in terms of free differential algebras. The elements of that 
algebra are usually higher-degree forms and they are additional degrees of freedom in the supergravity field 
spectrum besides the vielbein, the spin connection, and the gravitinos. 
There is a vast literature on the argument which we refer to for details and applications 
\cite{Chryssomalakos:1999xd,Bandos:2004xw,Bandos:2004ym,deAzcarraga:2005jd,Fre:2005px,Sati:2008eg,Fre:2008qw,Fiorenza:2010mh}

As discussed in several works, see for example \cite{if1,if2,if3,if4,if5},  it has been shown that there other sectors of cohomology for supergravity and supersymmetric theories, 
that play an important role in supergravity theory. It has been discovered, that in integral form cohomology 
(see \cite{Catenacci:2020ybi,Cremonini:2022vgz} for the precise definition) there are additional cohomology classes. They are expected because of the Hodge duality discovered in \cite{Castellani:2015ata,Castellani:2016ezd}. 
Those are cohomologies in the sector of integral forms and pseudoforms. In the presence of supermanifolds, 
the exterior bundle is not sufficient to describe the complete geometry and it has to be supplemented by the 
sector of integral forms. Those forms can be integrated into the supermanifold and can be explicitly constructed 
in terms of the delta function of commuting 1-forms and their derivatives. 
 Therefore, in the present work, we discuss whether the techniques developed in \cite{sully,book} 
can be adapted to this new framework, when integral form cohomologies are present and what it implies. 

It is shown that once the FDA for the superform sector has been constructed, using the ring structure 
of forms and module structure of integral forms, also the integral form sector is removed by suitable potentials. 
However, not completely. It is proven in sec. 1.2 that the potentials needed for superforms are not 
enough to compensate for all integral forms. Indeed, at least one requires at least a maximal picture to define 
the FIDA. There could happen that one integral form with the maximal-picture 
is not enough since the introduction of a maximal-picture potential might introduce new cohomology classes with higher 
pictures and for that one needs one more potential. 

What is the role of the maximal-picture potential (and occasionally also the additional one where 
new integral form cohomologies emerge)? There are two aspects to be discussed: 1) does it change the physical 
spectrum of the theory? 2) what is the role of this additional integral form? It is shown in sec. 3.3 that indeed, there are no 
additional degrees of freedom and the new integral form is related to the original spectrum of the theory. Concerning  
its role, we have to recall that to write action, we have to integrate over the entire supermanifold and 
this can be done with an integral form. The presence of a naive integral form in the spectrum indeed tells us that 
we have to build the integral form to construct a consistent action and its equations of motion \cite{if1,if2,if3,if4,if5}. One can relate 
this non-trivial integral form (or its potential) as a reflection of the existence of the Berezianian 
(see \cite{Noja:2021xos}) and the fact that there is a single non-trivial integral form seems to indicate that the Berezianian is a truly essential ingredient in the 
supergravity realm. 
 
 In sec. 1, we list and discuss the ingredients needed for the analysis and the general theory. 
 In sec. 2, we give some examples, starting from a toy example to a non-abelian coset manifold example. In sec. 3, we apply 
 the construction to D=4,6,11 models and we construct the complete FIDA using the Hodge duality. New cocycles are 
 shown and the relations among them are discussed. In sec. 4, we write some conclusions and open issues on some delicate mathematical questions we are not able to discuss in the present work.

\subsection{Free Differential Algebras (FDA)}

Given a Lie supergroup $\mathcal{G}$, we have a super Lie algebra which we denote by 
$\mathcal{L}_{G}$ with $n$ bosonic generators $T^{a}$ and $m$ fermionic generators $Q^{\alpha}$. Associated to each generator we introduce the dual Maurer-Cartan forms $(V^a, \psi^\alpha)$ 
satisfying the Maurer-Cartan equations 
\begin{eqnarray}
\label{FF_A}
\nabla V^a & =& f^a_{~~\alpha \beta} \psi^\alpha \wedge \psi^\beta+  f^a_{~~b c} V^b \wedge V^c\,, ~~~~~ 
\nonumber \\
\nabla \psi^\alpha &=&   f^a_{~~b \beta} V^b \wedge \psi^\beta \,. 
\end{eqnarray}
where $f^a_{~~\alpha \beta},  f^a_{~~\alpha \beta}$ and $f^a_{~~b \beta}$ are the structure constants 
satisfying the super Jacobi identities. The differential $d$ is the Chevalley-Eilenberg differential and 
it is nilpotent because of the Jacobi identities. If the bosonic generators $T^{a}$ corresponds 
to the translation generators on a supermanifold and $Q^{\alpha}$ the supersymmetry generators, 
then $(V^a, \psi^\alpha)$ represent the supervielbein associated to the manifold. If the supermanifold is 
seen as a coset supermanifold (for example $AdS$ superspaces such as $OSp(4|1)/SO(3,1)$) 
then the differential $\nabla$ becomes the covariant differential and one needs to introduce the spin connection associated with the subgroup $SO(3,1)$. The MC forms $(V^a, \psi^\alpha)$ carry form number equal to one and 
they are anticommuting and commuting, respectively, with respect to the wedge product 
\begin{eqnarray}
\label{FFB}
V^{a} \wedge V^{b} = - V^{b} \wedge V^{a}\,, ~~~~
V^{a} \wedge \psi^{\alpha} = \psi^{\alpha} \wedge V^{a}\,, ~~~~
\psi^{\alpha}\wedge  \psi^{\beta} = \psi^{\beta} \wedge \psi^{\alpha} \,.  
\end{eqnarray}
For convenience, we use the notation $E^{A} = (V^{a}, \psi^{\alpha})$ and collectively we denote by 
$f^{A}_{~BC}$ the structure constants where $A =(a,\alpha)$ runs over all indices.  

On the space of forms $\Omega^{(p)}(\mathcal{SM})$ with trivial coefficient (trivial module), 
one can compute the Chevalley-Eilenberg cohomology at every form number. For a general 
discussion on this point for super Lie algebras, we refer to \cite{Catenacci:2020ybi,Cremonini:2022vgz}. 
Once the cohomology classes are found, we have a set of new forms   ($I$ is a label to distinguish the 
different cocycles) 
\begin{eqnarray}
\Omega_{I}^{(p)} (\sigma)=\Omega_{I, A_1...A_p} E^{A_1} \dots E^{A_p},
\end{eqnarray}
such that 
\begin{eqnarray}
\nabla \Omega_I^{~(p)} \equiv d\Omega_I^{~(p)} + E^A \wedge D(T_A)_{I}^{~J} \Omega_J^{~(p)}=0,~~~~\Omega_{I}^{~(p)} \not= \nabla \Phi_{I}^{~(p-1)
}  \label{cohomology}
\end{eqnarray}
where the coefficient constant coefficient $\Omega_{I, ~A_1...A_p}$  and the index 
$i$ labels each class. For a non-trivial module, the coefficients $\Omega_{I, ~A_1...A_p}$ 
are not constants and transform in non-trivial representations of the original super Lie group. 
 
The possible {\sl free differential algebra} (FDA) extensions $\mathcal{L}'_G$ 
of a Lie algebra $\mathcal{L}_G$ have been studied in \cite{sully,gm3,gm12,book}, and rely 
on the existence of Chevalley-Eilenberg cohomology classes in $\Omega^{(p)}(\mathcal{SM})$ which 
is compensated by introducing new potentials $A_{I}^{(p)}$  to get 
\begin{eqnarray}\label{genFDA}
& & d E^A+\frac12 f^{A}_{~~BC} E^B \wedge E^C=0  \nonumber\\
& &  \nabla A_{I}^{(p-1)} + \Omega_{I}^{(p)} =0  \label{FDA1}
\end{eqnarray}
 It is clear that $\Omega_{I}^{(p)}$ differing by exact pieces $\nabla \Phi
_{I}^{(p-1)}$ lead to equivalent FDA's, via the redefinition $A_{I}^{(p-1)} \rightarrow A_{I}^{(p-1)} + \Phi_{I}^{(p-1)}$. 
The whole procedure can be repeated on the free differential algebra $\mathcal{L}'_G$ 
which now contains $E^A$, $A_{I}^{(p-1)}$. In terms of the new ingredients we have 
\begin{eqnarray}
\Omega_{I}^{(q)} = \Omega_{I, A_1...A_r}^{~~~~~~~~~~I_1...I_s} E
^{A_1} \wedge ... \wedge E^{A_r} \wedge A_{I_{1}}^{(p-1)} \wedge ... \wedge 
A_{I_{s}}^{(p-1)}
\label{polyFDA}\end{eqnarray}
satisfying the cohomology conditions \eqref{cohomology}. If such a polynomial 
exists, the FDA of eq.s {\eqref{FDA1} can be further extended to $\mathcal{L}''_G$. Of course, 
one would like to know if the procedure stops after a finite number of steps or if continues indefinitely. 
We will discuss in the next section a procedure to verify step-by-step if the complete FDA is built. 
This is based on the construction of the Hilbert-Poincar\'e series associated with the cohomology and 
then one can explicitly verify how the introduction of a new potential $A_{I}^{(p-1)}$ modifies 
the full cohomology. Once all classes for the FDA algebra have been discovered and suitably compensated by 
corresponding potentials, the final Hilbert-Poincar\'e series must be equal to 1. The Hilbert-Poincar\'e series 
will be reviewed in the next section, but the general procedure has been already been discussed in 
\cite{Cremonini:2022cdm}. 

As for ordinary Lie algebras, a dynamical theory based on FDA is obtained by 
introducing non-vanishing curvature for all forms of the FDA. 
This means, for example, that $d=11$ supergravity is based on a 
deformation of the fields $V,\omega,\psi,A, B$ such as the torsion $T^{a}$, 
the Riemann curvature $R^{ab}$, the gravitino curvature $\rho^{\alpha}$, 
the $A$- and $B$-curvature are different from zero.

The rewriting of the FDA's in terms of larger Lie (super)algebras,
by expressing the $p$-forms with $p > 1$ as products of 1-form fields involving 
new fields, has been considered already in \cite{DFd11} for $d=11$ supergravity. 
Recent developments of this idea can be found in \cite{FDAnew1,FDAnew2,FDAnew3}.

\subsection{Free Integro-Differential Algebras (FIDA)}

Working with a superalgebra, the MC 1-forms $E^A_{(1)}$ are either 
commuting or anticommuting 
$$E^A_{(1)} \wedge E^B_{(1)} = (-1)^{|A||B|} E^B_{(1)} \wedge E^A_{(1)}$$
depending on whether the indices $A,B$ refer to bosonic or fermionic generators. We distinguish them by setting $E^A=V^a$ in the case of bosonic generators $|A|=0$ and 
$E^A=\psi^\alpha$ (with |A|=1) 
in the case of fermionic generators. We are interested in the superform 
the sector of the theory, but also in the integral sector of the theory, and those are represented 
by expressions of the form
\begin{eqnarray}
\label{fillaA}
\Omega^{(p|q)} = \Omega^{(p|q)}_{[a_1 \dots a_r] (\alpha_{r+1}\dots \alpha_{p}) [\beta_{1} \dots \beta_q]} V^{a_1} \dots  V^{a_r} \psi^{\alpha_{r+1}} \dots  
\psi^{\alpha_{q}} \delta^{(g_1)} (\psi^{\beta_1}) \dots  
\delta^{(g_q)} (\psi^{\beta_q}) 
\end{eqnarray}
where we have taken into account also the Dirac delta's $\delta^{(g)}(\sigma^\beta)$ 
of the anticommuting MC forms $\psi^\alpha$. 
The index $(g)$ denotes the order of the derivative of the 
Dirac delta. We act on $\Omega^{(p|q)}$ with the usual differential and we use the 
distributional-like properties (the indices $\alpha$ and $\beta$ are not summed)
\begin{eqnarray}
\label{fillaB}
&&d \left( \delta^{(g)}(\psi^\beta)\right)  = \delta^{(g+1)}(\psi^\beta) d \psi^\beta\,, ~~~~~
\psi^{\beta} \delta^{(g)}(\psi^\beta) = -  \delta^{(g-1)}(\psi^\beta) \,, ~~~~\nonumber 
\\
&&\delta^{(g)}(\psi^\alpha) \delta^{(g')}(\psi^{\alpha'}) = -  \delta^{(g')}(\psi^{\alpha'}) 
\delta^{(g)}(\psi^\alpha) \,, ~~~~~ V^a \delta^{(g)}(\psi^\alpha) = - \delta^{(g)}(\psi^\alpha) V^a
\end{eqnarray}
The number of delta functions corresponds to the picture number and its range is 
from zero (superforms) to the fermionic dimension (integral forms). 
Notice that the differential $d$ changes the form number $p$, for $\Omega^{(p|q)}$, 
but it does not change the picture number $q$. For changing the latter, one needs a 
{\it Picture Changing Operator}  $\mathbb{Y}^{(0|1)}$ which is closed and not exact. 

Given two forms $\Omega^{(p|q)}$ and $\Omega^{(p'|q')}$, we can 
multiply them as follows 
\begin{eqnarray}
\label{fillaC} 
\Omega^{(p+p'|q+q')} =
\left\{
\begin{array}{cc}
  \Omega^{(p|q)}\wedge \Omega^{(p'|q')} & {\rm if} ~~ q+q' \leq m\\
0  & {\rm if} ~~ q+q' > m 
\end{array}
\right. 
\end{eqnarray}
Notice that due to \eqref{fillaB}, if the argument of two delta's in $  \Omega^{(p|q)}$ and $\Omega^{(p'|q')} $ is the same, their product 
vanishes (exactly in the same way as for two differential forms $V^a$). 

There are two possible types of PCOs, the raising PCO $\mathbb{Y}^{(0|1)}$, which rise the picture by 
one unit as follows, given $ \Omega^{(p|q)}$, we set 
\begin{eqnarray}
\label{fillaCA}
 \Omega^{(p|q)} \longrightarrow   \Omega^{(p|q+1)} = \mathbb{Y}^{(0|1)}\wedge \Omega^{(p|q)}
\end{eqnarray}
with the property $d \left( \mathbb{Y}^{(0|1)}\wedge \Omega^{(p|q)}\right)  = 
 \mathbb{Y}^{(0|1)} \wedge d  \Omega^{(p|q)}$. Notice that $ \Omega^{(p|q+1)} $ could vanish if 
 the Dirac delta in $ \mathbb{Y}^{(0|1)}$ is one of the Dirac delta in $\Omega^{(p|q)}$. The 
 maximal raising PCO is obtained by the wedge product of $ \mathbb{Y}^{(0|1)}$ along all possible 
 MC 1-forms $\psi^\alpha$. On the other side, we can construct the lowering PCO 
 as follows 
 \begin{eqnarray}
\label{fillaD}
 \Omega^{(p|q)} \longrightarrow  \Omega^{(p|q-1)} 
 =\mathbb{Z}^{(0|-1)} \Omega^{(p|q)} = \left[d,  \Theta\left( 
 \iota_{\hat X}\right)\right] \Omega^{(p|q)}
\end{eqnarray}
where $\hat X$ is an odd vector field and $\Theta\left(\iota_{\hat X}\right)$ can be conveniently defined using the integral representation of the Heaviside (step) function. If $\hat X = \hat X^{\alpha} \partial_{\alpha}$, then $\iota_{\hat X} \psi^{\alpha} = \hat X^{\alpha}$ and 
\begin{eqnarray}
\label{fillaDA}
\Theta(\iota_{\hat X}) = \lim_{\epsilon\rightarrow 0} \int_{-\infty}^{\infty} \frac{e^{- i t \iota_{\hat X}}}{t + i \epsilon} dt
\end{eqnarray}
We will come back to this PCO in the forthcoming sections. 
 
Let us for the moment consider the two extreme cases: superforms with picture number zero 
and integral forms with maximal picture number 
\begin{eqnarray}
\label{fillaE}
\Omega^{(p|0)} &=& 
\Omega^{(p|0)}_{[a_1 \dots a_r] (\alpha_{r+1}\dots \alpha_{p})} 
V^{a_1} \dots  V^{a_r} 
\psi^{\alpha_{r+1}} \dots  \psi^{\alpha_{p}} \nonumber \\
\Omega^{(p|m)} &=& 
\Omega^{(p|m), (\beta_{r+1}\dots \beta_{r})}_{[a_1 \dots a_{p+r}]}
V^{a_1} \dots  V^{a_{p+r}} \iota_{\beta_{1}} \dots  \iota_{\beta_{r}} \delta^m(\psi)
\end{eqnarray}
where $\delta^m(\psi)$ is the product of all delta's and $ \iota_{\beta} $ is 
the contraction of with respect to the odd vector field $D_\beta$. By using the Hodge 
star operator defined as in \cite{Castellani:2015ata,Castellani:2016ezd}, they are dual to each other 
\begin{eqnarray}
\label{fillaF}
\star \Omega^{(p|0)}  = \Omega^{(n-p|m)} 
\end{eqnarray}
and satisfy the following identity 
\begin{eqnarray}
\label{fillaG}
\Omega^{(p|0)} \wedge \star \Omega^{(p|0)}  = \prod_{a=1}^n V^a \prod_{\alpha=1}^m  \delta(\psi^\alpha) \equiv {\rm Vol}^{(n|m)}
\end{eqnarray}
and the right-hand side ${\rm Vol}^{(n|m)}$ represents 
the Berezinian top form on the space of MC forms. ${\rm Vol}^{(n|m)}$ is closed 
and not exact, therefore it serves as a volume form. If the superalgebra is 
a matrix superalgebra, ${\rm Vol}^{(n|m)}$ is just the superdeterminant. In paper 
\cite{Catenacci:2020ybi,Cremonini:2022vgz} it is proven that 
for each cohomology class (cocycles) in the superform sector $H^{(p|0)}$ 
there exists a corresponding integral form cocycle $H^{(n-p|m)}$. This 
corresponds to the usual Poincar\'e duality in the case of conventional manifolds.  

Now, according to the FDA techniques, for each cocycle of $ \omega^{(p|0)}_I \in H^{(p|0)}$, 
(where $I$ is a label for the cocycle) one introduces a new potential such that 
\begin{eqnarray}
\label{fillaH}
\nabla A^{(p-1|0)}_I =  \omega^{(p|0)}_I
\end{eqnarray} 
to trivialize the cohomology. This can be done for all cocycles in  $H^{(p|0)}$ 
leading to a set of potentials $A^{(p-1|0)}_I$. Those define a new set of differential forms 
which satisfies the generalized MC equation \eqref{genFDA}. Then, one has to compute the 
new extended cohomology with additional potentials to see whether new cocycles emerge
(an extended analysis for $D=4,6,10,11$ models with extended superspace has been 
performed in \cite{Cremonini:2022cdm}). The procedure can be iterated till we get empty cohomology and 
the complete free differential algebra is built $(V^a, \psi^\alpha, A^{(p-1)}_I)$.  

What happens on the integral form side? Suppose that the set of cocycles  $ \omega^{(p|0)}_I $ 
is finite-dimensional, this implies that also the set of potentials $A^{(p-1)}_I)$ 
is finite-dimensional. The set of cocycles is a ring 
\begin{eqnarray}
\label{fillaHA}
 \omega^{(p|0)}_I  \wedge  \omega^{(p'|0)}_J = 
 \sum_{K}  C_{IJ}^K \omega^{(p+p'|0)}_K 
\end{eqnarray}
with integer coefficients $C_{IJ}^K$. The ring structure translates on the 
potentials 
\begin{eqnarray}
\label{fillaHAA}
 A^{(p-1|0)}_I  \wedge d A^{(p'-1|0)}_J = 
 \sum_{K}  C_{IJ}^K A^{(p+p'-1|0)}_K \,. 
\end{eqnarray}
Now, we can compute the Hodge dual of each of them 
 \begin{eqnarray}
\label{fillaHB}
\star  \omega^{(p|0)}_I  =  \omega^{(n-p|m)}_I 
\end{eqnarray}
and due to theorem proved in \cite{Cremonini:2022cdm}, $\omega^{(n-p|m)}_I$, we can prove that 
$d \omega^{(n-p|m)}_I =0$ as follows. 
Suppose that $d \omega^{(n-p|m)}_I \neq 0$ and therefore 
\begin{eqnarray}
\label{HBA}
d \omega^{(n-p|m)}_I  = d (\star  \omega^{(p|0)}_I )= \Sigma^{(n-p+1|m)}_I
\end{eqnarray}
for a $(n-p+1|m)$ form $\Sigma^{(n-p+1|m)}_I$. We act again with the 
Hodge dual to get 
\begin{eqnarray}
\label{HBB}
d^\dagger  \omega^{(p|0)}_I =\star d \star  \omega^{(p|0)}_I =\Lambda^{(p-1|0)}_I
\end{eqnarray}
Together with the closure $d  \omega^{(p|0)}_I=0$ and the usual definition of the 
Laplace-Beltrami operator $\nabla = d d^{\dagger} + d^{\dagger} d$ 
we can write eq \eqref{HBB} as 
\begin{eqnarray}
\label{HBC}
\omega^{(p|0)}_I = d \left(\Delta^{-1} \Lambda^{(p-1|0)}_I \right)
\end{eqnarray}
where $\Delta^{-1}$ is a Green function of the Laplace-Beltrami differential 
$\Delta = d d^\dagger + d^\dagger d$. Therefore, if $ \Lambda^{(p-1|0)}_I$ is non vanishing, then $\omega^{(p|0)}_I$ 
is exact, which is in contradiction with the hypotheses that $\omega^{(p|0)}_I \in H^{(p|0)}$. 

The integral cocycles $ \omega^{(n-p|m)}_I $ do not form a ring, but 
a module with respect to the ring structure \eqref{fillaHA}
\begin{eqnarray}
\label{fillaHC}
\omega^{(p|0)}_I  \wedge \omega^{(p'|m)}_J = 
  \sum_{K}  D_{IJ}^K \omega^{(p+p'|m)}_K 
\end{eqnarray}
since the wedge product of a differential cocycle $ \omega^{(p|0)}_I$ 
with an integral cocycle $\omega^{(p'|m)}_J$ is still an integral cocycle 
and therefore it can be expanded on a basis with integer coefficients $D_{IJ}^K$. 
Notice that the form numbers are added, but the picture number does not change. 
Using the potentials $\Phi_I^{(p-1)}$ for the differential cocycles, 
we can write \eqref{fillaHC} as follows 
\begin{eqnarray}
\label{fillaHD}
d A^{(p-1|0)}_I  \wedge \omega^{(p'|m)}_J = d \left( 
A^{(p-1|0)}_I  \wedge \omega^{(p'|m)}_J \right) = 
  \sum_{K}  D_{IJ}^K \omega^{(p+p'|m)}_K 
\end{eqnarray}
which tells us that for non-vanishing coefficients $D_{IJ}^K$ 
the introduction of the potentials $A_I^{(p-1)}$ allows us to remove 
all integral cocycles except one! Indeed, the integral form 
$\omega^{(p|m)}_i$ with the lowest form number cannot be obtained by the 
module structure \eqref{fillaHC} and therefore the introduction of the 
potentials $A^{(p-1)}_I$ are not enough to construct the completely free 
differential algebra. Then, for the lowest form-number integral form, 
say $\omega^{(p_0|m)}_0$ we need a novel potential 
$A^{(p_0-1|m)}_0$ such that
\begin{eqnarray}
\label{fillaHE}
\omega^{(p_0|m)}_0 = \nabla A^{(p_0-1|m)}_0
\end{eqnarray}
Then, finally, the FIDA is 
spanned by the generators 
\begin{eqnarray}
\label{fillaHF}
\left( V^a, \psi^\alpha, A^{(p-1|0)}_I\,, A^{(p_0-1|m)}_0 \right) 
\end{eqnarray}
There are some remarks: 
\begin{enumerate}
\item Notice that among the integral cocycles, 
in the case of unimodular superalgebras, we always have the highest form 
integral cocycle 
\begin{eqnarray}
\label{fillaHG}
\star 1 = \omega^{(n|m)} =  {\rm Vol}^{(n|m)}
\end{eqnarray}
which is closed and not exact. This class becomes trivial once 
we have introduced the potentials $\Phi_I^{(p-1|0)}$ for the differential 
forms. Indeed, using \eqref{fillaG} 
we have 
\begin{eqnarray}
\label{fillaHH}
 {\rm Vol}^{(n|m)} = d \left( A^{(p-1|0)}_I \wedge \star \omega^{(p|0)}_I \right) = 
 d \left( A^{(p-1|0)}_I \wedge \omega^{(n-p|m)}_I \right)
\end{eqnarray}
 for any $I$ running over the set of independent cocycles. Then, having introduced all potentials one has left with two cocycles: the trivial constant $\omega_0^{(0|0)}$ 
 and the lowest integral form $\omega^{(p_0|m)}_0$. It may happen that $p_0 =0$, 
 and therefore $\omega^{(p_0|m)}_0 = \mathbb{Y}^{(0|m)}$, that is it coincides with 
 the product of all PCO's. Trivializing the latter, introduce an integral-form potential 
 such that $ \mathbb{Y}^{(0|m)} = d A^{(-1|m)}$ which can be used to check the consistency 
 between differential form and integral form cocycles.  
 \item There might be the possibility (see for example the $OSp(1|2)$ discussed in the 
 forthcoming section) that $\Phi^{(p_0-1|m)}_0$ is not enough to complete the FIDA. 
 Indeed, if $p_0 +m$ is even, the integral form $\omega^{(p_0|m)}_0$ is even and 
 its potential $A^{(p_0-1|m)}_0$ is odd. Therefore, the product
 \begin{eqnarray}
\label{fillaHI}
\omega^{(2p_0-1|2m)}_0 = A^{(p_0-1|m)}_0 \wedge \omega^{(p_0|m)}_0
\end{eqnarray}
is a cohomology class. Indeed, $d \omega^{(2p_0-1|2m)}_0 =  \omega^{(p_0|m)}_0
\wedge \omega^{(p_0|m)}_0 =0$ because of \eqref{fillaC}. Notice that we assumed that the 
picture carried by $A^{(p_0-1|m)}_0$ is different from that carried by $\omega^{(p_0|m)}_0$. Then, finally, 
we can remove $\omega^{(2p_0-1|2m)}_0$ by adding a further potential $A^{(2p_0-2|2m)}_0$. The role 
of those new potentials is not fully understood and it can be studied by analyzing its dynamics.  A similar role has been played by the $B$ field associated with the 7-cocycle in D=11 supergravity \cite{DFd11}. From the kinematical point of view, the $B$ field is introduced to construct the FDA but, it turns out to be dynamically irrelevant. 

\end{enumerate}

\subsection{Hilbert-Poincar\'e series}

Before moving to cocycle computations and their FDAs, we review the definition of Hilbert-Poincar\'e series and Poincar\'e polynomials. For $X$ a \emph{graded} vector space with direct decomposition into $p$-degree homogeneous subspaces given by $X = \bigoplus_{p \in \mathbb{Z}} X_p$ we call the formal series 
\begin{eqnarray}
\label{POIA}
\mathbb{P}_X(t) = \sum_p ({\rm dim} \, X_p) (-t)^p
\end{eqnarray}
the \emph{Hilbert-Poincar\'e series} of $X$. Notice that we have implicitly assumed that $X$ is a of \emph{finite type}, \emph{i.e.} its homogeneous subspaces $X_p$ are finite-dimensional for every $p.$ The unconventional sign in $(-t)^p$ takes into account the \emph{parity} of $X_p$, which takes values in $\mathbb{Z}_2$ and it is given by $p \, \mbox{mod}\, 2$: this will be particularly useful in supergravity where commuting and anticommuting 
variables are needed. If also $\dim X$ is finite, then $\mathbb{P}_X(t)$ becomes a polynomial $\mathbb{P}_X [t]$, called \emph{Poincar\'e polynomial} of $X$. The evaluation of the Poincar\'e polynomial at $t=1$ yields the so-called \emph{Euler characteristics} $\chi_{_X} = \mathbb{P}_X[t=1] = \sum_p (-1)^p \dim X_p$ of $X$. If we assume that the pair $(X, d)$ is a differential complex for $X$ a graded vector space and 
$d: X_p \rightarrow X_{p+1}$ for any $p$, then the cohomology $H_{d}^\bullet (X) = \bigoplus_{p\in \mathbb{Z}} H_{d}^p(X)$ is a graded space. 
if we denote $b_p (M) \equiv \dim H^p_{dR} (M)$ the $p$-th Betti number of $M$. The Poincar\'e polynomial of $M$, defined as (Euler-Poincar\'e formula) 
\begin{eqnarray}
\label{fiA}
\mathbb{P}_{M} [t] = \sum_{p} b_p (M) (-t)^p
\end{eqnarray}
is the generating function of the Betti numbers of $M$. This property is known as {\it telescopic nesting} which implies that from the easy computation of $\mathbb{P}_X(t)$, one deduces $\mathbb{P}_{H(X)}(t)$. From the latter 
one can read the cohomology classes by their gradings and the parity (see \cite{Cremonini:2022cdm}
 for a complete discussion on the subject and explicit computations).  

Even if the notion of Betti numbers is originally related to the topology of a certain manifold or topological space, by extension, in this paper we will call \emph{Betti numbers} the dimensions of any cohomology space valued in a field, in particular, we will call $p$-th Betti numbers of a certain Lie (super)algebra the dimension of its Chevalley-Eilenberg $p$-cohomology group $b_p (\mathfrak{g}) = \dim H^p_{CE} (\mathfrak{g}),$ so that the Hilbert-Poincar\'e series of the Lie (super)algebra $\mathfrak{g}$ is the generating function of its Betti number
\begin{eqnarray}
\label{fiB}
\mathbb{P}_{\mathfrak{g}} (t) = \sum_p b_p (\mathfrak{g}) (-t)^p.
\end{eqnarray}

Notice that we used the notation $\mathbb{P} (t)$ on purpose: indeed, as we shall see, the Chevalley-Eilenberg cohomology $H^\bullet_{CE} (\mathfrak{g})$ is not in general finite dimensional for a generic Lie superalgebra $\mathfrak{g}$. 
In our framework, it is convenient to introduce a second grading (picture number). 
In that case, the space is said to be 
{\it bigraded vector space} $X = \sum_{p,q \in \mathbb{Z}} X^{p,q}$, then the gradation 
$X = \sum_r X^{r}$ given by 
\begin{eqnarray}
\label{POID}
X^r = \sum_{p+q=r} X^{p,q}
\end{eqnarray}
is called the {\it induced total gradation}. One can write a {\it double Hilbert-Poincar\'e series} 
\begin{eqnarray}
\label{POIE}
\mathbb{P}_X(t,\tilde t) = \sum_{p,q} (-t)^p (-\tilde{t})^q {\rm dim}X^{p,q}
\end{eqnarray}
 which, in any case, allows easier identification of cohomological classes 
 (see, e.g., \cite{Catenacci:2020ybi, Cremonini:2022vgz} where double Hilbert-Poincar\'e series have been used to select different type cohomologies). In the forthcoming section, we use the second grading $\tilde t$ 
 to count the picture.

\section{Examples}

In this section, we analyze in detail the following examples: a toy model, an abelian-group manifold, 
a non-abelian group manifold, and a coset model. 
All these models are interesting examples of the theory discussed above. The charge/scale assignment is 
done to preserve the Maurer-Cartan equations, or equivalently, to commute with the differential $d$. 
In the abelian cases, the assignment is $t^{2}$ to bosonic 1-forms and $t$ to fermionic 1-forms, 
in non-abelian cases, the assignment corresponds to form number. We refer to 
\cite{Catenacci:2020ybi,Cremonini:2022cdm} for an extended and complete discussion 
on charge/scale assignments. 

\subsection{Toy Model Example}

We consider the bosonic $1$-forms $T$ and $b$ and the fermionic $1$-form $\psi$ with 
the Maurer-Cartan equations 
\begin{eqnarray}
\label{MC}
dt = - 2 T \wedge b + \psi^2\,, ~~~~~~~~ d\psi = \psi \wedge b\,, ~~~~~~~~
d b =0\,. 
\end{eqnarray}
(this example is taken from \cite{book} pag. 802, example n. 2). 
It is a solvable super-Lie algebra. 
We can proceed as follows: we introduce a $0$-form $\phi$, such that $b = d\phi$. 
Then we can rewrite the MC as follows\
\begin{eqnarray}
\label{MConeA}
T = e^{2 \phi} (dx + \theta d\theta)\,, ~~~~~~~~~
\psi = e^\phi d\theta\,, ~~~~~~~~~~
b = d \phi\,. 
\end{eqnarray}
we have introduced the new coordinates ($0$-forms) $x, \theta$ which describes the 
superline. The solvable super-Lie algebra can be thought of as the gauging 
of the scale invariance. The gauge field of the scale invariance is $\phi$. 

The cohomology is easily computed in terms of $x,\theta, \phi$. 
\begin{eqnarray}
\label{MConeB}
&\omega^{(0|0)} =1\,,  &\omega^{(1|0)} = b \,,  \nonumber \\
&\omega^{(0|1)} = (dx + \theta d\theta) \, \delta'(d\theta) \,, 
&\omega^{(1|1)} = d\phi\, \wedge (dx + \theta d\theta) \, \delta'(d\theta) \,,
\end{eqnarray}
but they can easily be reconverted in terms of the original 
\begin{eqnarray}
\label{MConeC}
\omega^{(0|0)} =1\,, ~~~~~~~~~~
\omega^{(1|0)} = b \,, ~~~~~~~~~~
\omega^{(0|1)} = T \, \delta'(\psi) \,, ~~~~~~~
\omega^{(1|1)} = b \wedge T \, \delta'(\psi) \,, ~~~~~~~
\end{eqnarray}
with the relations 
\begin{eqnarray}
\label{MCD}
\omega^{(0|0)}  \wedge \omega^{(1|1)} =  \omega^{(1|0)} \wedge \omega^{(0|1)}
\end{eqnarray}
Let us compute the Poincar\'e polynomial for the cohomology
\begin{eqnarray}
\label{MCDA}
\mathbb{P}(t, \tilde t) = (1 - t) + (t - t^2)\tilde t = (1-t) (1 + t \tilde t ) 
\end{eqnarray}
where the first equation has the following meaning: 
 $(1-t)$ is the cohomology of the superforms $\omega^{(0|0)}$ and $\omega^{(1|0)}$, 
 $(t -t^2)\tilde t$ is the cohomology of the integral forms $\omega^{(0|1)}$ and $\omega^{(1|1)}$. The 
 second equality means $(1-t)$ is the cohomology of $b$ times  $(1 + \tilde t ) $ which is 
 the Poincar\'e polynomial of CE cohomology of super translation on a super line 
 $(dx + \theta d\theta, d\theta)$ which scale as $t^2$ and $t$ \cite{Catenacci:2020ybi,Cremonini:2022vgz}.
 
As is well-known there is another important operator to be used, the rising PCO $Z$. This 
change the picture and it is written as $Z = [d, \Theta(\iota_D)]$ where $D$ is the vector 
field dual to $\psi$, namely $\iota_D \psi =1$. Then, we have 
\begin{eqnarray}
\label{MCE}
Z (\omega^{(0|1)}) = d \Big( \Theta(\iota_D)  T \, \delta'(\psi) \Big) = d  \Big(  \frac{T}{\psi^2} \Big) = 
   \frac{- 2 t\wedge b + \psi^2}{\psi^2} + 2 \frac{t\wedge b}{\psi^2} = 1\,, \nonumber \\
Z (\omega^{(1|1)}) = d \Big( \Theta(\iota_D) b \wedge T \, \delta'(\psi) \Big) = d  \Big(  \frac{b \wedge T}{\psi^2} \Big) = 
   \frac{b \wedge \psi^2}{\psi^2} = b\,,
\end{eqnarray}
so, it correctly maps cohomologies into cohomologies. Notice that it reduces the picture, but does not change 
the form number. In addition, we have to underline that in the intermediate step, we generated some 
{\it inverse} forms (a.k.a Large Hilbert Space, see for example \cite{Catenacci:2018xsv}) and they are converted into 
conventional superforms.   
 
Let us now discuss the FIDA. According to \cite{book}, in presence of 
new cohomology classes, one introduces new forms to cancel those cocycles. In particular, here 
we have two new forms, a $0$-from $b = d\phi$ and 
we introduce the $(-1|1)$ form $A^{(-1|1)}$ to cancel the cocycle $\omega^{(0|1)}$ as 
\begin{eqnarray}
\label{MCEA}
d A^{(-1|1)} =  \omega^{(0|1)} \,. 
\end{eqnarray}
This implies also
\begin{eqnarray}
\label{MCF}
d \left( - A^{(-1|1)}\wedge b \right) =   \omega^{(1|0)} \wedge \omega^{(0|1)} = b \wedge T\, \delta'(\psi) \,, 
\end{eqnarray}
We also have an additional interesting equation which is 
\begin{eqnarray}
\label{MCG}
Z( d A^{(-1|1)} ) = d Z(A^{(-1|1)} ) = Z(   \omega^{(0|1)}) = 1
\end{eqnarray}
The first equality follows from $[d, Z]=0$. The second equality is the definition of $d d A^{(-1|1)}$, 
the last equality is due to \eqref{MCE}. This implies 
\begin{eqnarray}
\label{MCH}
d Z(A^{(-1|1)} )  =1\,.
\end{eqnarray}
Namely, even the constants are not in the cohomology. 

Before proceeding, we still have to check whether there are other possible cocycles. 
We note that $A^{(-1|1)}$ has odd parity, (its d-variation is even), therefore there are 
no powers of $A^{(-1|1)}$. In addition, it carries picture +1. One could explore the possibility that 
this is a new picture, but eq. \eqref{MCE} seems to say that the picture is the same as of $\delta(\psi)$. 
This implies that we cannot consider combinations of the form $A^{(-1|1)} \wedge \delta(\psi)$. Since 
we have already explored all possibilities independent of $A^{(-1|1)}$, we are left with 
\begin{eqnarray}
\label{MCI}
\Omega = A^{(-1|1)}\wedge (\alpha(\psi) + \beta(\psi) b + \gamma(\psi) b + \rho(\psi) t\wedge b) 
\end{eqnarray}
where $\alpha(\psi), \dots, \rho(\psi)$ are polynomials of $\psi$. 
The closed forms turn out to be exact
\begin{eqnarray}
\label{MCL}
\Omega_C = \alpha'  A^{(-1|1)} \psi^2 + 2 \beta'  A^{(-1|1)} \psi^2 \wedge b = 
d \left( \alpha' A^{(-1|1)} e^{- 2 \phi} T + \beta' A^{(-1|1)} \psi^2 \right) 
\end{eqnarray}
where $\alpha', \beta'$ are numbers. 
This implies that there are no other cohomology classes in the present extended algebra. 
Thus, the free differential algebra is given by 
\begin{eqnarray}
\label{MCM}
T, \psi, \phi, A^{(-1|1)}
\end{eqnarray}
We can still consider $Z(A^{(-1|1)})$ as an inverse form and it automatically implies that 
even the constants are exact. 

\subsection{Abelian Group Manifold Example: U(1|1)}

Let us consider the explicit example of $\mathfrak{u}(1|1)$; the MC equations read
\begin{equation}\label{uA}
	dU = 0, \qquad dW = - \psi^+ \psi^-, \qquad d\psi^+ = U \psi^+, \qquad d \psi^- = -U \psi^- \ .
\end{equation}
The superform cohomology is given by
\begin{equation}\label{uB}
\omega^{(0|0)} = 1\,, ~~~~~~~
\omega^{(1|0)} = U\,, ~~~~~~
\omega^{(p|0)} = 0\,, p>1
\end{equation}
This means that only the abelian factor, whose associate MC form is $U$, is in the cohomology 
of superforms. The cohomology among integral forms read
\begin{equation}\label{uC}
	\omega^{(p|2)}=0, p\leq 1\,, ~~~~~~~
	\omega^{(1|2)} = W \delta(\psi^{+}) \delta(\psi^{-})\,, ~~~~~~
	\omega^{(2|2)} = U W \delta(\psi^{+}) \delta(\psi^{-})\,,
\end{equation}
the last expression corresponds to the Berezinian 
\begin{equation}\label{uD}
	\mathpzc{B}er = U \wedge W \wedge \delta \left( \psi^+ \right) \wedge \delta \left( \psi^- \right) \ .
\end{equation}

The two classes in $\omega^{(1|0)}$ and $\omega^{(1|2)}$ 
are dual, via the Berezinian complement duality 
\begin{equation}\label{uF}
	U \wedge \star U = \mathpzc{B}er \ ;
\end{equation}
notice that they live in two distinct (though quasi-isomorphic) complexes. In the present 
case the Hilbert-Poincar\'e polynomial is given by 
\begin{eqnarray}
\label{uFA}
\mathbb{P}(t, \tilde t) = (1 - t) (1 - t \tilde t^{2})
\end{eqnarray}
Each term corresponds to \eqref{uB} and \eqref{uC}, in particular we have $(1 - t)$ correspond to 
$\omega^{(0|0)}$ and $\omega^{(1|0)}$, while $(1- t) (-t \tilde t^{2}) = - t \tilde t^{2} + t^{2} \tilde t^{2}$ 
correspond to $\omega^{(1|2)}$ and $\omega^{(2|2)}$. 

Notice that by setting $\tilde t=1$ we recover the bosonic subgroup $U(1)\times U(1)$ and the 
Hilbert-Poincar\'e polynomial is just the product of the polynomial for each abelian factor. On the other 
side if we set $\tilde t =t$ we get the polynomial $(1- t)(1 - t^{3})$ which is the Poincar\'e polynomial 
of $U(2)$ and finally setting $\tilde t=0$ we get $(1-t)$ which is, according to Fuks theorem \cite{fuks},  just 
the superform cohomology and it corresponds to a $U(1)$.  

By following the constructive methods of FDAs (see, e.g., \cite{book}), we have to introduce new forms to the Lie superalgebra to trivialize the CE cohomology classes. In particular, in order to compensate for the class $\omega^{(1|0)} = U$, one has to introduce an even $(0|0)$-form $\omega_0$ such that 
\begin{equation}\label{TFDAA}
	d \omega_0 = U \ .
\end{equation}
We should now re-evaluate the CE cohomology of the FDA generated by $U,W,\psi^1,\psi^2,\omega_0$. Again, we will investigate both superforms and integral forms. It is not difficult to show that the only non-trivial cohomology group among superforms is $H^0$: the new closed objects introduced by $\omega_0$ are of the form
\begin{equation}\label{TFDAB}
	\omega^{(p+1|0)} = \omega_0 U \wedge \omega^{(p|0)} \left( \psi^+ , \psi^- \right) \ , \ \forall p \geq 0 \ ,
\end{equation}
where $\omega^{(p|0)} \left( \psi^+ , \psi^- \right)$ is a degree $p$ polynomial in the variables $\psi^+ , \psi^-$. If $p=0$ one has
\begin{equation}\label{TFDAC}
	\omega_0 U = \frac{1}{2} d \left[ \omega_0^2 \right] \ .
\end{equation}
If $p \geq 1$, by using the relations
\begin{eqnarray}\label{TFDAD}
	&&\omega_0 U \wedge \left( \psi^+ \right)^p = \frac{1}{p} d \left[ \omega_0 \wedge \left( \psi^+ \right)^p - \frac{1}{p} \left( \psi^+ \right)^p \right] \ ,\nonumber 
\\	&& \ \omega_0 \wedge U \wedge \left( \psi^- \right)^p = - \frac{1}{p} d \left[ \omega_0 \wedge \left( \psi^- \right)^p - \frac{1}{p} \left( \psi^- \right)^p \right] \ , \ \forall p \geq 1 \ ,
\end{eqnarray}
together with \eqref{uA} and \eqref{TFDAA}, it is easy to show that any superform as \eqref{TFDAB} is exact.
This can be easily checked by modifying the Hilbert-Poincar\'e polynomial: the introduction of the new form 
$\omega_{0}$ implies that 
\begin{eqnarray}
\label{TFDADA}
\mathbb{P}(t, \tilde t) \mapsto \frac{\mathbb{P}(t, \tilde t)}{(1-t)} = (1- t \tilde t^{2})
\end{eqnarray}

Let us move to integral forms: the first important remark involves the class $\mathpzc{B}er$. The introduction of the $(0|0)$-superform $\omega_0$ implies that $\mathpzc{B}er$ is exact:
\begin{equation}\label{TFDAE}
	\mathpzc{B}er = U \wedge W \wedge \delta \left( \psi^+ \right) \wedge \delta \left( \psi^- \right) = d \left[ \omega_0 W \wedge \delta \left( \psi^+ \right) \wedge \delta \left( \psi^- \right) \right] \ .
\end{equation}
This fact has as a direct implication the failure of the \virgolette Berezinian complement duality'' which then does not hold for the FDA $\mathfrak{g}'$. Analogously, notice that also the form $\mathpzc{B}er \otimes \omega_0^p , p \geq 0$, is exact:
\begin{equation}\label{TFDAF}
	\omega_0^p U \wedge W \wedge \delta \left( \psi^+ \right) \wedge \delta \left( \psi^- \right) = \frac{1}{p+1} d \left[ \omega_0^{p+1} W \wedge \delta \left( \psi^+ \right) \wedge \delta \left( \psi^- \right) \right] \ , \ \forall p \geq 0 \ .
\end{equation}
Notice that adding $\omega_{0}$ still defines a cohomology space, since it does not involve the form $U$ (hence, the introduction of $\omega_0$ does not spoil the non-exactness). This indicates that despite $U$ and $\iota_U \mathpzc{B}er$ being dual (in the sense of \eqref{uF}), they can not be compensated by a single new term in the FDA. Moreover, it is not difficult to prove that $\omega_0$ does not introduce new cohomology classes among integral forms and among superforms. Really, the new closed objects introduced by $\omega_0$ (except for \eqref{TFDAF}) are of the form
\begin{equation}\label{TFDAG}
	\omega^{(2-p|2)} = \omega_0 U \wedge \omega \left( \iota_+ , \iota_- \right) \delta \left( \psi^+ \right) \wedge \delta \left( \psi^- \right) \ ,
\end{equation}
where $\omega^{(-p|0)} \left( \iota_+ , \iota_- \right)$ is a (formal) degree $p$ polynomial in the variables $\iota_+ , \iota_-$. The exactness of terms as \eqref{TFDAG} is easily seen by considering \eqref{uA}, \eqref{TFDAA} and the relations
\begin{eqnarray}
	\label{TFDAH} \omega_0 U \wedge \iota_+^p \delta \left( \psi^+ \right) \wedge \delta \left( \psi^- \right) &=& - \frac{1}{p} d \left[ \omega_0 \iota_+^p \delta \left( \psi^+ \right) \wedge \delta \left( \psi^- \right) + \frac{1}{p} \iota_+^p \delta \left( \psi^+ \right) \wedge \delta \left( \psi^- \right) \right] \ , \nonumber \\
	\label{TFDAI} \omega_0 U \wedge \delta \left( \psi^+ \right) \wedge \iota_-^p \delta \left( \psi^- \right) &=& \frac{1}{p} d \left[ \omega_0 \delta \left( \psi^+ \right) \wedge \iota_-^p \delta \left( \psi^- \right) - \frac{1}{p} \delta \left( \psi^+ \right) \wedge \iota_-^p \delta \left( \psi^- \right) \right] \ .
\end{eqnarray}

We now want to trivialize the remaining cohomology class among integral forms. We introduce then a $(0|2)$-integral form $\eta^{0|2}$ so that
\begin{equation}
	d \eta^{0|2} = W \wedge \delta \left( \psi^+ \right) \wedge \delta \left( \psi^- \right) \ .
\end{equation}
This new generator does not introduce new cohomology classes. Since the new generator $\eta^{0|2}$ is 
commuting, we have to multiply the Hilbert-Poincar\'e polynomial with $1/(1- t \tilde t^{2})$ which cancels 
the remaining factor of \eqref{TFDADA}. In this way, we have completed the FIDA for this model.

\subsection{Non-Abelian Group Manifold Example: OSp(1|2)}

If we consider the super Lie-algebra $\mathfrak{osp}(1|2)$, there are 
the MC forms $V^a, \psi^\alpha$ with the MC equations 
\begin{eqnarray}
\label{CICA}
d V^a = \psi \gamma^a \psi + (V \wedge V)^a\,, ~~~~~~
d \psi^\alpha = V_a (\gamma^a\psi)^\alpha\,. 
\end{eqnarray}
We have shown that there are the following cocycles 
\begin{eqnarray}
\label{CICB}
&&\omega^{(0|0)} =1\,, ~~~~~~~~~~~~~
\omega^{(3|0)} =\frac12 \psi \gamma_a \psi V^a + \frac{1}{3!} (V\wedge V \wedge V)\,,\\
&&\omega^{(0|2)} = \frac12 (V\wedge V)^{ab} \iota \gamma_{ab} \iota \delta^2(\psi) + \delta^2(\psi)\,, ~~~~~~~
\omega^{(3|2)} = \frac{1}{3!} (V\wedge V \wedge V) \delta^2(\psi) \,. 
\end{eqnarray}
where we have displayed in the first line the superform cocycles and in the second line the 
integral form cocycles.  In addition, we have to recall that there are pseudoforms obtained by 
acting with $Z_{(\alpha)}$. There are two PCO $Z_{(\alpha)}$ associated to the two directions in the 
fermionic space. 
Then, we have 
\begin{eqnarray}
\label{CICC}
\omega^{(0|1)}_{(\alpha)} = Z_{(\alpha)} ( \omega^{(0|2)})\,, ~~~~~~~~~~
\omega^{(3|1)}_{(\alpha)} = Z_{(\alpha)} ( \omega^{(3|2)})\,, 
\end{eqnarray}
By consistency we have that $Z_{(1)} Z_{(2)}  \omega^{(0|2)} = \omega^{(0|0)}$ and 
$Z_{(1)} Z_{(2)}  \omega^{(3|2)} = \omega^{(3|0)}$. The complete set of cohomologies is easily 
described by the Hilbert-Poincar\'e polynomial 
\begin{eqnarray}
\label{CICD}
\mathbb{P}(t,\tilde t) = (1 - t^3) (1 + \tilde t)^2\,. 
\end{eqnarray}

Now, we can construct the FDA for the present model. At first, we introduce a 2-form $B^{(2|0)}$ 
to cancel the cocycle $\omega^{(3|0)}$. This has a twofold effect, 
it cancels $ \omega^{(3|0)}$, but it also cancels $ \omega^{(3|2)}$ 
as follows 
\begin{eqnarray}
\label{CICCA}
 \omega^{(3|2)} = d \left( B^{(2|0)} \wedge \omega^{(0|2)} \right) 
\end{eqnarray}
We still have one additional cocycle to cancel: $ \omega^{(0|2)}$. 
This can be done by introducing the $-1$-form potential with 
picture number $+2$ 
as 
\begin{eqnarray}
\label{CICE}
d A^{(-1|2)} = \omega^{(0|2)}
\end{eqnarray}
to compensate the cocycle $\omega^{(0|2)}$ (which is the PCO $\mathbb{Y}$). 
Notice that the ring/module structure implies also 
\begin{eqnarray}
\label{CICF}
d \Big( A^{(-1|2)} \wedge \omega^{(3|0)}  \Big) =  \omega^{(3|2)} 
\end{eqnarray}
which means again that the volume form $\omega^{(3|2)}$ is written in terms of the 
additional $A^{(-1|2)}$ and the superform $\omega^{(3|0)}$. This is consistent 
with eq. \eqref{CICCA} since $A^{(-1|2)} \wedge \omega^{(3|0)} = B^{(2|0)} \wedge \omega^{(0|2)}  + d ({\rm exact})$. 
As discussed in sec. 3.1, we have still one cocycle of the form 
\begin{eqnarray}
\label{CICFA}
\omega^{(-1|4)} = A^{(-1|2)} \wedge \omega^{(0|2)}
\end{eqnarray}
which requires a new potential $B^{(-2|4)}$ to complete the FIDA. Still,
at the moment we do not have a physical interpretation of those new ingredients, 
and they play a role algebraically. 

In addition, we can 
act with the PCO $Z_{(\alpha)}$ on $A^{(-1|2)}$ to get 
\begin{eqnarray}
\label{CICG}
d \left[ Z_{(\alpha)} \Big( A^{(-1|2)} \Big) \right] = \omega^{(0|1)}_{(\alpha)} \nonumber \\
\end{eqnarray}
namely, the new forms 
\begin{eqnarray}
\label{CICH}
 A^{(-1|1)}_{(\alpha)} = Z_{(\alpha)} \Big( A^{(-1|2)} \Big)\,, ~~~~~~~~~~~~
 \end{eqnarray}
cancel the cocycle of the pseudo-form type, but they are not new ones.  
So, we conclude by observing that the FIDA is 
generated by 
\begin{eqnarray}
\label{CICK}
V^a\,,~~ \psi^\alpha\,,~~ B^{(2|0)}\,, ~~~~ A^{(-1|2)},~~ \,, B^{(-2|4)}\,,
\end{eqnarray}

A remark: 
Notice there is an additional form $A^{(-1|0)} = Z_{(1)} Z_{(2)} \Big( A^{(-1|2)} \Big)$ 
which has an interesting behaviour of $ A^{(-1|0)}_{(\alpha)}$: 
$d \left[ Z_{(1)} Z_{(2)} \Big( A^{(-1|2)} \Big) \right] = \omega^{(0|0)} =1$
which implies that in the present case, even the constants are exact forms. 
This is in a complete analogy with the superstring Hilbert space: 
the cohomology (vertex operators) of the BRST superstring charge selects the physical states 
on the small Hilbert space, but in the Large Hilbert Space (where the zero mode $\xi_0$ is 
introduced) the BRST cohomology is empty. Not even the constants are considered cohomology 
classes, if one admits inverse forms. Notice that, once we have $A^{(-1|0)}$, every closed form can be made exact. In addition, we 
should observe that $A^{(-1|0)}$ is an inverse form and this is a completely new ingredient in the 
framework. Its role has to be understood in this new framework.

\subsection{Coset Manifold Example: OSp(1|4)/SO(1,3)}

We  consider the case with $D=4$ and $N=1$ described by 
the supercoset manifold $OSp(1|4)/SO(1,3)$ 
It has four bosonic dimensions and four fermionic dimensions. The vielbeins $V^a$ are obtained by the natural identification of Maurer-Cartan forms of $\mathfrak{osp}(1|4)$ 
\begin{eqnarray}
\label{MCA}
V^{\alpha \beta} = \gamma_a^{\alpha\beta}V^a + \gamma_{ab}^{\alpha\beta} \omega^{ab}
\end{eqnarray}
in terms of Dirac matrix decomposition 
where $\omega^{ab}$ is the spin connection of $SO(1,3)$. 
The vielbeins and the spin connection satisfy the MC equations 
\begin{eqnarray}
\label{MCAA}
R^{ab} &\equiv& d \omega^{ab} + \omega^{ac} \wedge \omega_{c}^{~b} = V^a\wedge V^b + 
\frac{1}{2} \bar\psi \gamma^{ab} \psi  \,, \nonumber \\
\rho &\equiv& d \psi  + \frac14 \omega^{ab} \gamma_{ab} \psi  =
 \frac{i}{2} V^a \gamma_a \psi \,,\nonumber\\
T^a &\equiv& d V^a  + \omega^{a}_{~b} V^b  = \frac{i}{2}\bar\psi \gamma^a \psi \,. 
\end{eqnarray}
The supervielbeins $\psi^\alpha$ are 
in the Majorana representation of $SO(1,3)$. 

The study the cohomology classes of the coset $OSp(1|4)/SO(1,3)$ can be easily done 
knowing the Chevalley-Eilenberg classes of $OSp(1|4)$ and those of $SO(1,3)$ (using the 
Hochshild-Serre spectral sequence). 
We list them in terms of Poincar\'e polynomials 
\begin{eqnarray}
\label{MCB}
\mathbb{P}_{OSp(1|4)} (t, \tilde t) &=& (1 - t^3)(1- t^7) (1+ \tilde t^4) \nonumber \\
\mathbb{P}_{SO(1,3)}(t) &=& (1 - t^3)^2 \nonumber \\
\mathbb{P}_{OSp(1|4)/SO(1,3)} (t, \tilde t) &=& \frac{(1 - t^4)(1- t^8) (1+ \tilde t^4) }{(1 - t^4)^2} = (1+t^4) (1+ \tilde t^4)
\end{eqnarray}
where $t$ counts the form degree and $\tilde t$ counts the picture degree. For the 
last equation we have used the theorem by Greub-Vanstone-Halperin which allows 
us to write the Poincar\'e polynomial for coset spaces, but also we have studied at each picture number the different cohomologies. 

Notice that in the case of the 
supergroup $OSp(1|4)$, the factor $(1+\tilde t^4)$ indicates that there are two sets of classes: the superforms and 
the integral forms. Explicitly we have the four classes 
\begin{eqnarray}
\label{MCC}
&\omega^{(0|0)} = 1\,, ~~~~~~
&\omega^{(4|0)} = \bar\psi \gamma_{ab} \psi V^a V^b + \epsilon_{abcd} V^a V^b V^c V^d\,, 
\nonumber \\
&\omega^{(0|4)} = \delta^4(\psi) + V^a V^b \bar\iota \gamma_{ab} \iota  \delta^4(\psi)\,, ~~~~~
&\omega^{(4|4)} = \epsilon_{abcd} V^a  V^b V^c V^d  \delta^4(\psi) \,. 
\end{eqnarray}
with the relation $\omega^{(0|0)} \wedge \omega^{(4|4)}  = \omega^{(4|0)} \wedge \omega^{(0|4)}$. They are all closed and not exact. 
In particular the class $\omega^{(0|4)}$ is to be identified with the PCO which can be used to compute the action in superspace (see \cite{inprep}). 
To simplify the description of the result it is convenient to use 
a chiral/antichiral decomposition 
and we make the cosmological constant $\lambda$ explicit 
\begin{eqnarray}
\label{MCAAB}
R^{\alpha\beta} &\equiv& d \omega^{\alpha\beta} + \omega^{\alpha\beta} \epsilon_{\gamma\delta}  \omega^{\delta \beta} = \lambda^2 V_2^{\alpha\beta} + 
\lambda \psi^{\alpha} \psi^\beta  \,, \nonumber \\
R^{\dot\alpha\dot\beta} &\equiv& d \omega^{\dot\alpha\dot\beta} + 
\omega^{\dot\alpha\dot\beta} \epsilon_{\dot\gamma\dot\delta}  \omega^{\dot\delta\dot \beta} =  \lambda^2 V_2^{\dot\alpha\dot\beta} + 
 \lambda \bar\psi^{\dot\alpha} \bar\psi^{\dot\beta}  \,, \nonumber \\
\rho^\alpha &\equiv& d \psi^\alpha  + \frac14 \omega^{\alpha\beta} \epsilon_{\beta\gamma}
\psi^\gamma  =   \lambda V^{\alpha\dot\alpha} \epsilon_{\dot\alpha \dot\beta} \bar\psi^{\dot\beta}\,,\nonumber\\
\rho^{\dot\alpha} &\equiv& d \bar\psi^{\dot\alpha}  + \frac14 \omega^{\dot\alpha\dot\beta} \epsilon_{\dot\beta\dot\gamma}
\bar\psi^{\dot\gamma}  =  \lambda V^{\alpha\dot\alpha} \epsilon_{\alpha \beta} \psi^{\beta}\,,\nonumber\\
T^{\alpha\dot\alpha} &\equiv& d V^{\alpha\dot\alpha}  + 
\omega^{\alpha\gamma}\epsilon_{\gamma\delta} V^{\delta\dot\alpha}  + 
\omega^{\dot\alpha\dot\gamma}\epsilon_{\dot\gamma\dot\delta} V^{\dot\delta\alpha} =  \bar\psi^{\dot\alpha}  \psi^\alpha \,. 
\end{eqnarray}
by setting $\lambda =0$, we reproduce the flat MC equations and the 
Lorentz subgroup decouples from the MC equations (actually it plays a role as a semidirect product of $\mathfrak{siso}(1,3)$ of the super-Poincar\'e). 
Notice that the spin connection $\omega^{ab}$ is split into a self-dual and anti-self dual part $\omega^{\alpha\beta} , \omega^{\dot\alpha\dot\beta}$. In the same way, we have the 
self-dual and anti-self dual curvatures $R^{\alpha\beta} , R^{\dot\alpha\dot\beta}$ and 
the combinations $V_2^{\alpha\beta}$ and $V_2^{\dot\alpha\dot\beta}$. 
Using the cosmological constant, we can redefine the scaling of the different 
coordinates as $[V]=t^2, [\psi]=t, [\omega]=t^0$ and $[\lambda] = t^{-2}$. This scaling commutes with eqs. \eqref{MCAAB}. 
Now we can express the cohomology classes on a new basis as follows 
\begin{eqnarray}
\label{newbA}
\omega^{(0|0)} &=& 1\,, 
~~~~~~~~~~~~~~~~~~~~~~~~~~~~~~~~~~~~~~~~~~~~~~~~~~~~~~~~~~~~\longleftrightarrow ~~ 1 
\nonumber \\
\omega^{(4|0)} &=&  \left(
V_2^{\alpha\beta} \psi_\alpha \psi_\beta + 
V_2^{\dot\alpha\dot\beta} \bar\psi_{\dot\alpha} \bar\psi_{\dot\beta}\right) 
 +  \lambda V^4\,, 
~~~~~~~~~~~~~~~~~~~ \longleftrightarrow ~~ t^6
\nonumber\\
\omega^{(0|4)} &=& \lambda^{-1} \delta^4(\psi) + 
 \left( V_2^{\alpha\beta} \iota_\alpha \iota_\beta \delta^4(\psi)+ 
V_2^{\dot\alpha\dot\beta} \bar\iota_{\dot\alpha} \bar\iota_{\dot\beta} \delta^4(\psi) \right) \,,  ~~ \longleftrightarrow ~~ t^2  \tilde t^4 \nonumber \\
\omega^{(4|4)} &=& V^4 \delta^4(\psi)\,. ~~~~~~~~~~~~~~~~~~~~~~~~~~~~~~~~~~~~~~~~~~~~~~~~~~\longleftrightarrow ~~ t^8 \tilde t^4
\end{eqnarray}
The constant $\lambda$ is placed in such a way as to respect the 
ring structure $\omega^{(4|0)}  \wedge \omega^{(0|4)} = \omega^{(4|4)}$. 
Notice that compared with the flat case, only 2 classes of each sector (superforms, 
and integral forms) survive the curvature of the space. In particular, the cocycle 
$\omega^{(3)}_2$ discussed above disappears from the cohomology. This is 
consistent with the fact that the cohomology discussed is the 
equivariant Chevalley-Eilenberg cohomology of the coset space.  

In the present case, the FIDA is easily built. We have to add 
an anticommuting $(3|0)$-form $A^{(3|0)}$ which scales as $t^6$ and 
an anticommuting $(0|4)$-form $A^{(-1|4)}$ which scales as $t^2 \tilde t^4$ 
such that 
\begin{eqnarray}
\label{newbB}
d A^{(3|0)} = \omega^{(4|0)} \,, ~~~~~
d A^{(-1|4)} = \omega^{(0|4)}
\end{eqnarray}
and in turn, we also have to add two commuting potentials $B^{(2|0)}$ and 
$B^{(-2|4)}$  which scale as $t^{8}$ and $t^4 \tilde t^8$. 
This renders the cohomology trivial as can be seen by using the Poincar\'e polynomial 
\begin{eqnarray}
\label{newbC}
P^{FIDA}_{OSp(1/4)/SO(1,3)} = (1 + t^6) (1 + t^2 \tilde t^4) \frac{(1- t^6)}{(1-t^{12})} 
\frac{(1- t^2 \tilde t^4)}{(1-\tilde t^4 t^{8})} = 1
\end{eqnarray}
The complete analysis of psuedoforms will be deferred to future publications. 

\section{Hodge Dual Operator, Dual Cocycles and Harmonic Cocycles}

\subsection{D=4}
The computation for the cocycles has been discussed and performed in \cite{Cremonini:2022cdm}, 
we use here the same notations and conventions. 

Let's use the Hodge dual construction: given a $(p|0)$-form in the superspace $\mathbb{R}^{(4|4)}$, 
$\omega^{(p)}(V, \psi, \bar\psi)$ we can calculate its Hodge dual by the formula 
\begin{eqnarray}
\label{HO1A}
\star \omega^{(p)}(V, \psi, \bar\psi) = \# \int e^{i ( 
V^{\alpha\beta} \epsilon_{\alpha\beta} \epsilon_{\dot\alpha\dot\beta} \sigma^{\beta\dot\beta} + 
 \psi^\alpha \epsilon_{\alpha\beta} b^\beta   + \bar\psi^{\dot\alpha} \epsilon_{\dot\alpha\dot\beta} \bar b^{\dot\beta}) } 
  \omega^{(p)}(\sigma, b, \bar b) [d^4\sigma d^2 b d^2 \bar b]
\end{eqnarray}
where $\sigma, b, \bar b$ are auxiliary variables needed to define the Hodge dual operation on a given form. 
The coefficient  $\#$ is computed by choosing the signature of the superspace, requiring the 
idempotency of the star operation (see \cite{Castellani:2015ata,Castellani:2016ezd}) 
and it is irrelevant for the 
present purposes. 
Then, we have 
\begin{eqnarray}
\label{HO1B}
\omega^{(4|4)}_0 = \star 1 &=  V^4  \delta^4(\psi) \,, 
&~~\longleftrightarrow  ~~t^8 \tilde t^4
\nonumber \\ 
\omega^{(1|4)}_1 = \star \omega^{(3)}_1  &=  V^{\alpha\dot\alpha}_3 \iota_{\alpha} \bar \iota_{\dot\alpha} \delta^4(\psi) \,, 
&~~\longleftrightarrow~~ - t^4 \tilde t^4
\nonumber \\ 
\omega^{(0|4)}_2 = \star \omega^{(4)}_2 &= V^{\alpha\beta}_2 \iota_{\alpha} \iota_{\beta} \delta^4(\psi) \,, 
&~~\longleftrightarrow~~  t^2 \tilde t^4
\nonumber \\ 
\omega^{(0|4)}_3 = \star \omega^{(4)}_3 &= V^{\dot\alpha\dot\beta}_2 \bar\iota_{\dot\alpha} 
\bar\iota_{\dot\beta} \delta^4(\psi) \,,  
&~~\longleftrightarrow~~  t^2 \tilde t^4 \nonumber \\ 
\omega^{(-1|4)}_4 = \star  \omega^{(5)}_4 &= V^{\alpha\dot\alpha} \iota_{\alpha} \bar \iota_{\dot\alpha} \delta^4(\psi)\,, 
&~~\longleftrightarrow ~~ - \tilde t^4 \nonumber \\
\omega^{(0|4)}_5 = \star  \omega^{(4)}_5 &= \delta^4(\psi) \,,  
&~~\longleftrightarrow ~~ \tilde t^4 \,,
\end{eqnarray}
where $\iota_\alpha$ and $\bar\iota_{\dot\alpha}$ are the derivatives of the Dirac delta's  $\delta^4(\psi)$, with 
respect to their arguments. The scale $\tilde t$ is assigned to every single delta $\delta(\psi)$. Notice that all forms are closed except $\omega^{(-1|4)}_4$ 
which gives 
\begin{eqnarray}
\label{HO1C}
d \omega^{(-1|4)}_4 = \psi^\alpha \bar \psi^{\dot \alpha} \iota_{\alpha} \bar \iota_{\dot\alpha} \delta^4(\psi) = 4\,  \delta^4(\psi) = 4\, \omega^{(0|4)}_5
\end{eqnarray}
by integration by parts. Therefore, cohomology in that sector is represented by the Poincar\'e polynomial 
\begin{eqnarray}
\label{HO1D}
\mathbb{P}_{N=1}(t) = (2 t^2 - t^4 + t^8) \tilde t^4
\end{eqnarray}
which is exactly the Poincar\'e dual expression to 

\begin{eqnarray}
\label{HnewA}
\mathbb{P}_{N=1}(t) = (1 - t^4 + 2 t^6) 
\end{eqnarray} 
Notice that, differently from the usual cohomologies, the Poincar\'e duality cannot be 
established in the same complex of differential forms, but it has to be searched into the complex of integral forms. The overall $\tilde t^4$ stands for the picture equal to 4 of each term.  

The complete Poincar\'e polynomial has the following form 
\begin{eqnarray}
\label{HnewB}
\mathbb{P}_{N=1}(t) = (1 - t^4 + 2 t^6) + (2 t^2 - t^4 + t^8) \tilde t^4
\end{eqnarray}

One might also consider a partial Hodge dualization. As 
follows 
\begin{eqnarray}
\label{HO1E}
\star_C \omega^{(p)}(V, \psi, \bar\psi) = \# \int e^{i ( 
V^{\alpha\beta} \epsilon_{\alpha\beta} \epsilon_{\dot\alpha\dot\beta} \sigma^{\beta\dot\beta} + 
 \psi^\alpha \epsilon_{\alpha\beta} b^\beta ) } 
  \omega^{(p)}(\sigma, b, \bar \psi) [d^4\sigma d^2 b]
\end{eqnarray}
where only the dual of $\psi$ is considered. $\star_C$ stands for chiral Hodge dual. 

In that case, we have the following 
expressions 
\begin{eqnarray}
\label{HO1F}
\omega^{(4|2)}_0 = \star_C 1 &=  V^4  \delta^2(\psi) \,, 
&~~\longleftrightarrow  ~~t^8 \tilde t^2
\nonumber \\ 
\omega^{(3|2)}_1 = \star_C \omega^{(3)}_1  &=  V^{\alpha\dot\alpha}_3 \iota_{\alpha} \bar \psi_{\dot\alpha} \delta^2(\psi) \,, 
&~~\longleftrightarrow~~ - t^6 \tilde t^2
\nonumber \\ 
\omega^{(0|2)}_2 = \star_C \omega^{(4)}_2 &= V^{\alpha\beta}_2 \iota_{\alpha} \iota_{\beta} \delta^2(\psi) \,, 
&~~\longleftrightarrow~~  t^2 \tilde t^2
\nonumber \\ 
\omega^{(4|2)}_3 = \star_C \omega^{(4)}_3 &= V^{\dot\alpha\dot\beta}_2 \bar\psi_{\dot\alpha} 
\bar\psi_{\dot\beta} \delta^2(\psi) \,,  
&~~\longleftrightarrow~~  t^6 \tilde t^2 \nonumber \\ 
\omega^{(1|2)}_4 = \star_C  \omega^{(5)}_4 &= V^{\alpha\dot\alpha} \iota_{\alpha} \bar \psi_{\dot\alpha} \delta^2(\psi)\,, 
&~~\longleftrightarrow ~~ - t^2 \tilde t^2 \nonumber \\
\omega^{(0|2)}_5 = \star_C  \omega^{(4)}_5 &= \delta^2(\psi) \,,  
&~~\longleftrightarrow ~~ \tilde t^2 \,,
\end{eqnarray}
Notice that we have computed the different cocycles with respect only the 
variables $V$ and $\psi$. Accordingly, we have computed the form degree and 
the picture number. 

By computing the closure of the different expressions 
we get 
\begin{eqnarray}
\label{HO1G}
d \omega^{(4|2)}_0 &=& 0 \,, 
\nonumber \\ 
d \omega^{(3|2)}_1  &=&  \omega^{(4|2)}_3  \,, \nonumber \\ 
d \omega^{(0|2)}_2  &=&  \omega^{(1|2)}_4  \,, \nonumber \\ 
d \omega^{(4|2)}_3  &=& 0 \,,  
 \nonumber \\ 
d \omega^{(1|2)}_4  &=& 0\,, 
\nonumber \\
d \omega^{(0|2)}_5  &=& 0 \,,  
\end{eqnarray}
and this implies that there are only two cohomology classes $\omega^{(4|2)}_0$ 
and $\omega^{(0|2)}_5$ (the chiral volume form and the chiral PCO). 
This can be expressed in terms of the Poincar\'e polynomial 
as follows 
\begin{eqnarray}
\label{HO1GA}
\mathbb{P}_{N=1}(t) = 2  (1+ t^8) \tilde t^2
\end{eqnarray}
where factor 2 stands for the chiral and the antichiral representations.

\subsection{D=6}
The computation for the cocycles has been discussed and performed in \cite{Cremonini:2022cdm}, 
we use here the same notations and conventions. 

In this section, we introduce the Hodge dual operator for the flat supermanifold $(6|16)$ 
underlying the model $N=(4,0)$. We recall the supervielbeins $(V_{\alpha\beta}, \psi^A_\alpha)$ are respectively 
anticommuting and commuting 1-forms and we introduce the Hodge dual operator $\star$. 
For that, we need a metric on the supermanifold space 
$(\eta^{[\alpha\beta] [\delta\gamma]}, \eta^{\alpha \beta}_{AB})$ to construct scalar products, 
then one need dual variables $(\sigma_{\alpha\beta}, b^A_\alpha)$ (in the same representation of 
$(V_{\alpha\beta}, \psi^A_\alpha)$) and finally, given a form $\omega(V, \psi)$ we 
set
\begin{eqnarray}
\label{hodA}
\star \omega(V, \psi) = \# \int \omega(\sigma, b) e^{i (\sigma_{\alpha\beta} 
\eta^{[\alpha\beta] [\delta\gamma]} V_{\delta\gamma} +   b^A_\alpha  \eta^{\alpha \beta}_{AB} \psi^B_\beta)} 
[d^6\sigma d^{16}b] 
\end{eqnarray}
where $  \omega(\sigma, b)$ is the original form $\omega(V, \psi)$ where we substitute the supervielbeins in terms 
of the dual variables. The coefficient $\#$ is needed to implement the idempotency: $\star^2 =1$. 

In the vectorial representation (since it is an $SO(6)$ representation) we have $\eta^{ab}$ using 
the vector indices $a,b=1,\dots, 6$, or written in the spinorial representation (antisymmetric tensor 
of $SU(4)$) it reads $\eta^{[\alpha\beta] [\delta\gamma]} = \epsilon^{\alpha\beta \delta\gamma}$. 
In the spinorial representation for $\psi^A_\alpha$ however, there is no such metric. This is 
due to the well-known properties of the $SU(4)$ group. Therefore,  $\eta^{\alpha \beta}_{AB})$ does 
not exist. This means the Hodge dual operation $\star$ is not invertible and therefore it is not well defined. 
However, we can restrict the action of the Hodge dual operator on the bilinear expressions in the 
spinors, where we can define a well-defined action of the Hodge dual operation. 

Note that since the $R$-symmetry is $Usp(4)$ for $N=(4,0)$ and $SU(2)$ for $N=(2,0)$, the only invariant expressions are built with the antisymmetric tensors $\mathbb C^{AB}$ or $\epsilon^{ABCD}$ (in the case of $USp(4)$). Since only the spinor fields $\psi_\alpha^A$ carry those indices 
we have that the bilinear expressions are always antisymmetric in the spinorial indices $\alpha, \beta, \dots$. Therefore, we can set the Hodge dual operation 
of those bilinears as follows (that can be also deduced by an integral formula as above by introducing dual variables $b_{\alpha\beta\gamma}$) 
\begin{eqnarray}
 \label{hoB}
 \star \left( \psi^{[A}_{[\alpha} \psi^{B]}_{\beta]}\right)  = V^6 
 \epsilon_{\alpha\beta\gamma\delta} \mathbb C^{AR} \mathbb C^{BS}
 \iota^\gamma_R \iota^\delta_S \delta^{(16)}(\psi) 
\end{eqnarray}
which is a $(4|16)$ integral form. Notice that the contraction $\iota^\alpha_A = \partial/\partial \psi^A_\alpha$ removes one-degree form and it carries the opposite representation with respect to $\psi^A_\alpha$. The Hodge dual is defined such that 
\begin{eqnarray}
 \label{hoC}
 \psi^{[R}_{[\gamma} \psi^{S]}_{\delta]} \wedge \star \left( \psi^{[A}_{[\alpha} \psi^{B]}_{\beta]}\right)  =  
 \epsilon_{\alpha\beta\gamma\delta} \mathbb C^{AR} \mathbb C^{BS}
V^6 \delta^{(16)}(\psi) 
\end{eqnarray}
where the integral form $\omega^{(6|16)}_0 = V^6 \delta^{(16)}(\psi)$ 
is the Berezinian of the supermanifold. Decomposing the supervielbeins along the 
curved coordinates $(x^m, \theta^A_\mu)$ of the supermanifold we have 
\begin{eqnarray}
\label{berA}
V^a = E^a_m dx^m + E^{a, \mu}_{A} d\theta^{A}_{\mu}\,, ~~~~~
\psi^A_\alpha = E^A_{\alpha,m} dx^m + E^{A, \mu}_{\alpha, B}d\theta^{B}_{\mu}\,, 
\end{eqnarray}
 where $(E^a_m, E^{a,\mu}_{A}, E^A_{\alpha,m}, E^{A, \mu}_{\alpha, B})$ 
 are superfields in the coordinates $(x^m, \theta^A_\mu)$ and
 \begin{eqnarray}
\label{berB}
 \omega^{(6|16)}_0 = V^6 \delta^{(16)}(\psi)={\rm Sdet} \left( \begin{array}{cc} 
   E^a_m    &   E^{a, \mu}_{A}\\
   E^A_{\alpha,m}   & E^{A, \mu}_{\alpha, B}
 \end{array}\right) d^6x \delta^{(16)}(d\theta)\,. 
 \end{eqnarray}
The matrix in the superdeterminant is a supermatrix with dimension 
$(6 + 16) \times (6+16)$. 
So, the definition of the Hodge dual corresponds to the definition of the 
Berezinian complement, given a superform $\omega^{(p)}$, its Hodge dual 
is an integral form $\omega^{(6-p|16)} = \star \omega^{(p)}$ such that
\begin{eqnarray}
 \label{berC}
 \omega^{(p)}\wedge  \star \omega^{(p)} = F(x^m, \theta^A_\mu) 
 V^6 \delta^{(16)}(\psi)
\end{eqnarray}
where $F(x^m, \theta^A_\mu) = ||\omega^{(p)}||^2 $.   

Let us now consider the case $N=(2,0)$ and 
$N=(4,0)$ with the classes $\omega^{(4)}_1,  \omega^{(3)}_2, \omega^{(6)}_3$ and $\omega^{(7)}_4 $
computed in \cite{Cremonini:2022cdm}, their 
Hodge duals read
\begin{eqnarray}
 \label{newTA}
 \omega^{(2|16)}_1 = \star\omega^{(4)}_1 &=& \frac{1}{6!}V^6 \epsilon^{ABCD} \eta^{ab} 
 \bar\iota_{[A} \gamma_a \iota_{B]} \, 
 \bar\iota_{[C} \gamma_b \iota_{D]} \delta^{16}(\psi) \,, \nonumber \\
\omega^{(3|16)}_2 = \star \, \omega^{(3)}_2 &=& \frac{1}{5!}
\epsilon^{abcdef} 
 V_b V_c V_d V_e V_f
\mathbb C^{AB} 
 \bar\iota_{[A} \gamma_a \iota_{B]} \delta^{16}(\psi)
 \,, \nonumber \\
 \omega^{(0|16)}_3 =
\star\, \omega^{(6)}_3 &=& 
\delta^{16}(\psi)
\,, \nonumber \\ 
\omega^{(-1|16)}_4 = \star \, 
 \omega^{(7)}_4 &=& V^a 
 \mathbb C^{AB} 
 \bar\iota_{[A} \gamma_a \psi_{B]}\delta^{16}(\psi) \,, 
 \nonumber \\
 \omega^{(6|16)}_0 = \star \omega^{(0)}_0 &=& V^6 \delta^{16}(\psi)\,. 
\end{eqnarray}
The indices over the new integral forms denote the 
form degree and the picture number. Now, we can apply the differential operator $d$ and we use the MC equations to get 
\begin{eqnarray}
 \label{newTB}
 &d \omega^{(2|16)}_1 = \omega^{(3|16)}_2\,, ~~~~
 &d  \omega^{(3|16)}_2 =0 \,, \nonumber \\
 &d \omega^{(-1|16)}_4 = \omega^{(0|16)}_3 \,,  
 ~~~~
 &d \omega^{(0|16)}_3 =0\,. 
\end{eqnarray}
In the case of $N=(2,0)$, the invariant $\omega^{(2|16)}_1 =0$ and the invariant 
$\omega^{(3|16)}_2$ is a cohomology class in the 
integral form sector, together with  
the class $\omega^{(6|16)}_0$ given 
in \eqref{berA}. 

The polynomial $\mathbb{P}_{N=(2,0)}$ 
is 
\begin{eqnarray}
 \label{newTC}
\mathbb{P}_{N=(2,0)} = - (1 - t^4) t^8 \tilde t^{16}
\end{eqnarray}
where $\tilde t^{16}$ counts the picture number of 
the integral forms. The term $t^{12} \tilde t^{16}$ 
corresponds to the Berezinian $\omega^{(6|16)}_0$ 
while $- t^{8} \tilde t^{16}$ corresponds to the 
class $\omega^{(3|16)}_2$. The signs respect the statistic of the cohomology classes. 

Finally, notice that acting with Hodge dual on 
\eqref{newTB} we had 
\begin{eqnarray}
 \label{newTCA}
 & \star \, d \star \omega^{(4)}_1 = \omega^{(3)}_2\,, ~~~~
 &  \star \, d \star \omega^{(3)}_2 =0 \,, \nonumber \\
 &\star \, d \star \omega^{(7)}_4 = 
 \omega^{(3)}_3 \,,  
 ~~~~
 &\star\, d \star \omega^{(3)}_3 =0\,. 
\end{eqnarray}
Combining these equations with the closure of the cocycles, and using the conventional Laplace-Beltrami 
$\Delta = d d^\dagger + d^\dagger d$ and $d^\dagger = \star d \star$
we get 
\begin{eqnarray}
\label{newTD}
\Delta \omega^{(4)}_1 &=& \omega^{(4)}_1 \,, ~~~~~ 
\Delta \omega^{(3)}_2 = \omega^{(3)}_2 \,, ~~~~~ \nonumber \\
\Delta \omega^{(6)}_3 &=& \omega^{(7)}_4 \,, ~~~~~
\Delta \omega^{(7)}_4 = \omega^{(7)}_4 \,, 
\end{eqnarray}
In the case $N=(2,0)$, since $\omega^{(4)}_1 =0$, 
there are two cohomology classes 
$\omega^{3}_2$ and $\omega^{(3|16)}_3$ and they 
satisfy
\begin{eqnarray}
\label{newTE}
\Delta \omega^{(3)}_2 =0\,, ~~~~
\Delta \omega^{(3|16)}_3 =0\,. 
\end{eqnarray}
This shows that the cocycles which are cohomology classes are also harmonic in the usual sense, 
while those classes which are not in the cohomology are eigenforms 
of the Lapalce-Beltrami operator with a non-zero eigenvalue. This corresponds to the Hodge theorem for 
superforms.

In the present context, in the case of $N=(2,0)$ and 
$N=(2,2)$ we can apply the procedure to identify new forms to 
add to the theory. Again, we use the conventions and the results of  \cite{Cremonini:2022cdm}. 

In the case of $N=(2,0)$, we have the form cocycles 
$\omega^{(0)}_0, \omega^{(3)}_2, \omega^{(6)}_3, \omega^{(7)}_4$ and the integral cocycles 
$\omega^{(3|16)}_2, \omega^{(0|16)}_3, 
\omega^{(-1|16)}_4, \omega^{(6|16)}_0$. Computing the cohomology 
we are left with 
\begin{eqnarray}
 \label{FDAA}
 \omega^{(0)}_0, \omega^{(3)}_2\,, ~~~~~~~~
 \omega^{(3|16)}_2\,,  \omega^{(6|16)}_0
\end{eqnarray}
represented by the polynomial (putting together forms and 
integral forms) 
$$\mathbb{P}_{N=(2,0)}(t) = (1- t^4)(1 - t^8 \tilde t^{16})$$
Now, to follow to FDA technique, we cancel the $ \omega^{(3)}_2$ 
cocycle by adding the $A^{(2)}$ form (which scales as $t^4$, according to our conventions) such that 
\begin{eqnarray}
 \label{FDAB}
 d A^{(2)} = \omega^{(3)}_2\,, 
\end{eqnarray}
Notice that the newborn $A^{(2)}$ carries no representation of the 
R-symmetry and Lorentz group, it is a $2$-form and therefore it is 
a commuting field with respect to the wedge product. 
Notice that we can form wedge products between forms and integral forms and we can immediately observe that 
\begin{eqnarray}
 \label{FDABA}
 d \left( A^{(2)} \wedge \omega^{(3|16)}_2\right)   = \omega^{(6|16)}_0\,, 
\end{eqnarray}
namely, also the cohomology class $\omega^{(6|16)}_0$ is trivialized! This is rather striking: the Berezinian class 
$\omega^{(6|16)}_0$ becomes an exact form and it drops out from the cohomology. Notice that also the scales match correctly: $ A^{(2)} \wedge \omega^{(3|16)}_2$ scales as $t^{12} \tilde t^{16}$ as 
the Berezinian $\omega^{(6|16)}_0$. 

Did we completely trivialize the cohomology? Let us check it 
by using the Poincar\'e polynomial. To compute the polynomial 
for the FDA we just divide $\mathbb{P}_{N=(2,0)}(t)$, by the 
contribution of $A^{(2)}$, namely 
\begin{eqnarray}
 \label{FDAC}
 \mathbb{P}_{N=(2,0)}(t) \longrightarrow 
 \mathbb{P}_{N=(2,0)}^{FDA}(t)= \frac{(1- t^4)(1 - t^8 \tilde t^{16})}{(1-t^4)} = 
 (1 - t^8 \tilde t^{16})
\end{eqnarray}
which clearly shows that the cohomology related to $(1-t^4)$ is 
removed, but there is still a remainder. Indeed, we still have 
one cohomology class around: $\omega^{(3|16)}_2$. Let us 
follow the FDA prescription and introduce the 
integral form $A^{(2|16)}$ with scales $t^8 \tilde t^{16}$ 
such that 
\begin{eqnarray}
 \label{FDAD}
 d A^{(2|16)} = \omega^{(3|16)}_2\,, 
\end{eqnarray}
which finally cancels the last cohomology class. Notice 
that it seems the natural ingredient: it is a $2$-integral form to be compared with the conventional $2$-form introduced 
in \eqref{FDAB}, but they are not related. This is crucial since 
it appears that the new FDA requires new ingredients never seen before.  Finally, observe that 
$A^{(2|16)}$ is again commuting, invariant tensor and therefore 
we can apply the computation of the Poincar\'e polynomial as above
\begin{eqnarray}
 \label{FDAE}
 \mathbb{P}_{N=(2,0)}^{FDA}(t) \longrightarrow 
 \mathbb{P}_{N=(2,0)}^{FIDA}(t)= \frac{(1 - t^8 \tilde t^{16})}{1 - t^8 \tilde t^{16})} = 1
\end{eqnarray}
which signifies that we have trivial cohomology and we 
were able to construct a complete and consistent FDA. 
Notice that it is also true
\begin{eqnarray}
 \label{FDAF}
 d \left( A^{(2|16)}\wedge \omega^{(3)}_2 \right)  = \omega^{(3|16)}_2\,, 
\end{eqnarray}
but that implies that $\left( A^{(2|16)}\wedge \omega^{(3)}_2 \right)$ differs from $\left( A^{(2)} \wedge \omega^{(3|16)}_2\right)$ by exact terms, indeed we have that 
\begin{eqnarray}
 \label{FDAG}
 A^{(2|16)}\wedge \omega^{(3)}_2  + 
  A^{(2)} \wedge \omega^{(3|16)}_2
 = d  \left( A^{(2)} \wedge A^{(2|16)}_2 \right) 
\end{eqnarray}
as a consistency check. 

Two important remarks: {\it 1)} Notice that 
we have treated $A^{(2)}$ as a commuting quantity. Indeed, 
we have multiplies the Poincar\'e polynomial by $1/(1-t^4)$. 
The expansion of which leads to 
\begin{eqnarray}
 \label{FDAGA}
\frac{1}{1- t^4} = 1 + t^4 + t^8 + t^{12} + \dots \,, ~~
\longleftrightarrow 1\,, A^{(2)}\,, A^{(2)}\wedge A^{(2)}\,, 
A^{(2)}\wedge A^{(2)}\wedge A^{(2)}\, \dots.....
\end{eqnarray}
which implies that we have to take any power of $A^{(2)}$. 
That can be understood if we admit that this form is expanded 
on a basis of superforms that allow any power of them. {\it 2)} 
what about $A^{(2|16)}$? Again, we have adopted the pragmatic point of view: we have considered as a commuting quantity and 
therefore we admit any power of. Nevertheless, this clashes with the picture number. It carries picture number 16 and therefore, we cannot admit any more power of it. We can conceive a way out by introducing new commuting spinorial $1$-form $\eta$ in the game as in Fr\'e-D'Auria 
algebra and the related pictures $\delta(\eta)$ \cite{DFd11,FDAnew1,FDAnew2}.  

\subsection{D=11}

The computation of cocycles for 11d supergravity can be done using the Molien-Weyl formula 
and it gives
\begin{eqnarray}
\label{11dG}
\mathbb{P}_{11d}(t) = 1 + t^6
\end{eqnarray}
where the $t^6$ stands for $\omega^{(4)} = \bar\psi \Gamma_{ab} \psi V^a V^b$. There is only 
one cocycle in the present sector and the result is consistent with the literature \cite{DFd11}. 
To construct the FDA, we have to add a 3-form $A^{(3)}$ which scales with $t^6$ such that 
\begin{eqnarray}
\label{11dH}
d A^{(3)} = \omega^{(4)}
\end{eqnarray}
Then the resulting Poincar\'e polynomial becomes 
\begin{eqnarray}
\label{11dGA}
 \mathbb{P}_{11d}(t,t^2) = 1 + t^6 \longrightarrow P^{FDA}_{11d}(t,t^2) =  (1 + t^6)(1 - t^6) = 1 - t^{12} 
\end{eqnarray} 
This means that the FDA is not complete. Indeed, we see immediately, that there is 
a new cohomology class 
\begin{eqnarray}
\label{11dL}
\omega^{(7)} = A^{(3)} \wedge \omega^{(4)}_3 - \bar \psi \Gamma_{a_1 \dots a_5} \psi V^{a_1} \dots V^{a_5}
\end{eqnarray}
To cancel that class, one needs to introduce a further commuting potential $B^{(6)}$ such that 
$d B^{(6)} = \omega^{(7)}$. Therefore, the final expression for the Poincar\'e polynomial 
is 
\begin{eqnarray}
\label{11dGB}
P^{FDA}_{11d}(t,t^2) =  (1 + t^6)(1 - t^6) = 1 - t^{12} \longrightarrow P^{FDA}_{11d}(t,t^2) =  \frac{(1 + t^6)(1 - t^6)}{1- t^{12}} = 1  
\end{eqnarray} 
Notice that in this expression, the factor $1/(1-t^{12})$ takes into account the infinite series of the powers 
$(B^{(6)})^k$. 

The relevant Fierz identity is 
\begin{eqnarray}
\label{curC}
(\bar\psi \Gamma^{ab} \psi) (\bar\psi \Gamma_a \psi) =0\,.  
\end{eqnarray}
and bi-spinor decomposition is 
\begin{eqnarray}
\label{curD}
\psi {}_\wedge \bar \psi = 
\frac{1}{32} \left( \Gamma_a (\bar \psi \Gamma^a \psi) - \frac12 \Gamma_{a b} 
(\bar \psi \Gamma^{ab} \psi) + \frac{1}{5!}  
\Gamma_{a_1 \dots a_5} (\bar \psi \Gamma^{a_1 \dots a_5} \psi) \right) 
\end{eqnarray}

We recall that there are two interesting forms written in terms of 1-forms $\psi^\alpha$ and vielbeins $V^a$: 
\begin{eqnarray}
\label{cicA}
\omega^{(4)} = \bar\psi \Gamma_{ab} \psi V^a V^b\,, ~~~~~~~~~
\omega^{(7)} = \bar\psi \Gamma_{a_1 \dots a_5} \psi V^{a_1} \dots V^{a_5}\,, 
\end{eqnarray}
They satisfy the following equations 
\begin{eqnarray}
    \label{cicB}
    d \omega^{(4)} =0\,, ~~~~~
    d \omega^{(7)} = -\frac12 \omega^{(4)}_3\wedge \omega^{(4)}_3\,.
\end{eqnarray}
The second equation is a consequence of the Fiersz identities 
$$\bar\psi \Gamma_{a_1 \dots a_5} \psi \bar\psi \Gamma^{a_5} \psi = 
\bar\psi \Gamma_{[a_1 a_2} \psi \bar\psi \Gamma_{a_3 a_4]} \psi$$ 
Let us consider the Hodge dual of those superforms
\begin{eqnarray}
    \label{cicC}
    \omega^{(7|32)}=\star \omega^{(4)} &=& V^{a_1} \dots V^{a_9} \epsilon_{a_1\dots a_9 b_1b_2}\,  \bar\iota \Gamma^{b_1 b_2}\iota \delta^{32}(\psi)\,, \nonumber \\
    \omega^{(4|32)}=\star \omega^{(7)} &=& V^{a_1} \dots V^{a_6} \epsilon_{a_1\dots a_6 b_1 \dots b_5}\, \bar\iota \Gamma^{b_1 \dots b_5}\iota \delta^{32}(\psi)\,, ~~~~
\end{eqnarray}
where $\bar\iota \Gamma^{b_1 b_2}\iota = 
\frac{\delta}{\delta \bar{\psi}}\Gamma^{b_1 b_2}\frac{\delta}{\delta{\psi}} $ are the derivatives with respect to the argument of the delta functions. 
Therefore they act by integration by parts. In particular, if we compute the wedge product of $\omega_4$ with $\star\omega_4$ (and analogously for $\omega_7$) 
we get the volume form
\begin{eqnarray}
\label{cicCA}
\omega^{(4)} \wedge \star \omega^{(4)} = V_1 \dots V_{11} \delta(\psi_1) \dots \delta(\psi_{32})\,, ~~~~
\omega^{(7)} \wedge \star \omega^{(7)} = V_1 \dots V_{11} \delta(\psi_1) \dots \delta(\psi_{32})\,. 
\end{eqnarray}

Notice that the first one has degrees $(7|32)$ (due to the presence of $9$ vielbeins and two derivatives), while the second one has degree $(4|32)$. Notice that both are closed 
\begin{eqnarray}
\label{cicD}
d \star \omega^{(4)} &=& 9\,  (\bar\psi \Gamma^{a_1} \psi)  
V^{a_2} \dots V^{a_9} \epsilon_{a_1\dots a_9 b_1b_2}\,  \bar\iota \Gamma^{b_1 b_2}\iota \delta^{32}(\psi) \nonumber \\
&=& 
9 \, {\rm tr}(\Gamma^{a_1} \Gamma^{b_1 b_2})  
V^{a_2} \dots V^{a_9} \epsilon_{a_1\dots a_9 b_1b_2}\, \delta^{32}(\psi) =0\,, \nonumber \\
d \star \omega^{(7)} &=&  6\, 
V^{a_2} \dots V^{a_6} \epsilon_{a_1\dots a_6 b_1 \dots b_5}\, \bar\iota \Gamma^{b_1 \dots b_5}\iota \delta^{32}(\psi) \nonumber \\
&=& 6 \, {\rm tr}(\Gamma^{a_1} \Gamma^{b_1 \dots b_5})  
V^{a_2} \dots V^{a_9} \epsilon_{a_1\dots a_6 b_1 \dots b_5}\, \delta^{32}(\psi) = 0
\end{eqnarray}
they vanish because of the trace between the gamma matrices. On the other hand, 
if we compute the Hodge dual of $d\omega^{(7)}$, we get 
\begin{eqnarray}
\label{cicE}
\star d \omega^{(7)} = - \frac14 V^{a_1} \dots V^{a_7} \epsilon_{a_1 \dots a_7 b_1 \dots b_4}  
\bar\iota \Gamma^{b_1 b_2}\iota  \bar\iota \Gamma^{b_3 b_4}\iota \delta^{32}(\psi) 
\end{eqnarray}
Using the Fiersz identities, we can recast the derivatives 
as follows
\begin{eqnarray}
\label{cicEA}
\star d \omega^{(7)} = - \frac14 V^{a_1} \dots V^{a_7} \epsilon_{a_1 \dots a_7 b_1 \dots b_4}  
\bar\iota \Gamma^{b_1 b_2 b_3 b_4 b_5}\iota  \bar\iota \Gamma_{b_5}\iota \delta^{32}(\psi) 
\end{eqnarray}
and then we can compute the differential
\begin{eqnarray}
\label{cicF}
d \star d \omega^{(7)} = - \frac74 \bar\psi \Gamma^{a_1} \psi \dots V^{a_7} \epsilon_{a_1 \dots a_7 b_1 \dots b_4}  
\bar\iota \Gamma^{b_1 b_2 b_3 b_4 b_5}\iota  \bar\iota \Gamma_{b_5}\iota \delta^{32}(\psi) 
\end{eqnarray}
The integration by part produces two different structures: one vanishes because of the usual trace of gamma 
matrices, but the second structure gives the 
expression 
\begin{eqnarray}
\label{cicG}
d \star d \omega^{(7)}  = - \frac72 \star \omega^{(7)} ~~~~~ \Longrightarrow ~~~~\star d \star d \omega^{(7)} = - \frac72 \omega^{(7)}\, 
\end{eqnarray}
then finally it leads (together the vanishing of $d \star \omega_7 =0$, 
to the equations 
\begin{eqnarray}
\label{cicH}
\Delta \omega^{(7)} = - \frac72 \omega^{(7)}\,,  ~~~~~~ 
\Delta \omega^{(4)}_3 = 0\,. 
\end{eqnarray}
The second equation follows from $d \omega_4=0$. Then, we found that those forms satisfy a (massive) Laplace-Beltrami equation(as discussed in \cite{Grassi:2023yfe}, this is an indication that 
a Hodge theory may be established for 
supermanifolds). 

One question arises. Can one compute the dual forms from the Molien-Weyl formula? For that, we need to change 
the plethystic exponential for $\psi$ as follows (for details see \cite{Cremonini:2022cdm})
\begin{eqnarray}
\label{duald11A}
PE\left[\frac{1}{t} \chi_{32}(z_1, \dots, z_5)\right] = \tilde t^{32} \prod_{i=1}^{32} \frac{1}{(1 - \chi_{32, i} 1/t)} \,, ~~~~~~
\end{eqnarray}
where the parameter $t$ counts the degree of forms. This is to 
count the number of derivatives $\iota_{D_i}$ (which technically is 
the contraction along an odd vector field $D_i$) acting on $\bigwedge_{i=1}^{32}\delta(\psi^i)$. The 
factor $\tilde t^{32}$ counts the picture number. 
Thus, we finally get the result
\begin{eqnarray}
\label{duald11B}
\mathbb{P}_{\star 11d} (1/t, u) = 
\frac{(1-u) \left(u \left(1-\frac{u^4}{t^6}-\frac{u^6+u^4+u^2}{t^4}-\frac{u^8+u^4+1}{t^2}+\left(u^2+u+1\right) \left(u^6+u^3+1\right) u+1\right)\right)}{\frac{1}{1-t^4}}\nonumber 
\end{eqnarray}
which leads to 
\begin{eqnarray}
\label{duald11BA}
\mathbb{P}_{\star 11d} (1/t, t^2) = - (1+ t^{6}) t^{16} \tilde t^{32}  = - (t^{16} + t^{22}) \tilde t^{32} 
\end{eqnarray}
which represent the two classes
\begin{eqnarray}
\label{dual11C}
\omega^{(7|32)} &=& \star\omega^{(4)} =  V^{a_1} \dots V^{a_9} \epsilon_{a_1\dots a_9 b_1b_2}\,  \bar\iota \Gamma^{b_1 b_2}\iota \delta^{32}(\psi)\nonumber \\
\omega^{(11|32)} &=& \star 1 =  V^{1} \dots V^{11}  \delta^{32}(\psi)\,.
\end{eqnarray}
The overall sign in \eqref{duald11B} correctly detects the parity of those classes. 
Note that it does not appear in the cohomology the dual form $\star \omega^{(7)}$. Indeed, 
it can be easily shown that this form is exact 
\begin{eqnarray}
\label{dual11D}
\star \omega^{(7)} &=& d \omega^{(3|32)}\,, \nonumber \\
\omega^{(3|32)} &=& 
 V^{a_1} \dots V^{a_7} \epsilon_{a_1\dots a_7 b_1 \dot b_4}\, \bar\iota \Gamma^{b_1 b_2}\iota  \bar\iota \Gamma^{b_3 b_4}\iota\delta^{32}(\psi)\
\end{eqnarray}

Then, finally, we can compute the FDA of 11d superspace in both sectors: superforms
and integral forms. The complete Poincar\'e polynomial is 
\begin{eqnarray}
\label{totA}
\mathbb{P}_{11d} (t) =(1+t^6) - (1+ t^{6}) t^{16} \tilde t^{32}  =(1+t^6) (1 -  t^{16} \tilde t^{32})  
\end{eqnarray}
Therefore, following the prescription of the FDA, we have to add 
the 3-form $A^{(3)}_6$ and the 6-form $B^{(6)}_{12}$  such that 
\begin{eqnarray}
\label{totB}
d A^{(3)}_6 &=& \omega^{(4)}_6 \nonumber \\
d B^{(6)}_{12} &=& \omega^{(7)}_{12} - A^{(3)}_6 \wedge  \omega^{(4)}_6 
\end{eqnarray}
which implies, at the level of the Poincar\'e polynomial 
\begin{eqnarray}
\label{totC}
\mathbb{P}_{11d} (t) =(1+t^6) (1 -  t^{16} \tilde t^{32})   \longrightarrow 
 P^{FDA}_{11d} (t) = (1 -  t^{16} \tilde t^{32})  
\end{eqnarray}
Therefore we need to compensate the last cocycle 
by adding an {\it integral} potential, we add the $(6|32)$ 
form such that 
\begin{eqnarray}
\label{totD}
d B^{(6|32)} = \omega^{(7|32)} 
\end{eqnarray}
Then, the spectrum of the FDA is given by $A^{(3)}, B^{(6)}, B^{(6|32)}$. 
\vskip .5cm
As can be read in paper \cite{DFd11}, the $B^{(6)}$ does not modify 
the Lagrangian, since grouping all terms leads to an exact expression and 
therefore this does not affect the equations of motion. On the other hand 
the   $B^{(6|32)}$ might jeopardize the interpretation introducing new dof's. 
However, if we set 
\begin{eqnarray}
\label{totE}
F^{(4)} = d A^{(3)} = \star F^{(7|32)} = \star d B^{(6|32)}
\end{eqnarray}
 we finally related the dof's of the four form $F^{(4)}$ of the CJS supergravity 
 with those of the new integral potential $B^{(6|32)}$. The question of whether this equation 
 can be derived by an action principle is not known at the moment, but the existence of 
 pseudoforms in picture 16 might be crucial. 
To verify \eqref{totE} we list the components of $F^{(4)}$ 
\begin{eqnarray}
\label{totF}
F^{(4)} = F^{(4)}_{a_1 \dots a_4} V^{a_1}\dots V^{a_4} + 
 F^{(4)}_{a_1 \dots a_3 \alpha_1} V^{a_1}\dots V^{a_3} \psi^{\alpha_1} + \dots 
 + F^{(4)}_{
\alpha_1 \dots \alpha_4} \psi^{\alpha_1}\dots \psi^{\alpha_4} 
\end{eqnarray}
while for the dual form 
\begin{eqnarray}
\label{totG}
F^{(7|32)} = \left( {\mathcal F}_{a_1 \dots a_7} V^{a_1}\dots V^{a_7} + 
 {\mathcal F}_{a_1 \dots a_8}^{\alpha_1} V^{a_1}\dots V^{a_8} \iota^{\alpha_1} + \dots 
 + {\mathcal F}^{\alpha_1 \dots \alpha_4}_{a_1 \dots a_{11}} V^{a_1} \dots V^{a_{11}} 
 \iota^{\alpha_1}\dots \iota^{\alpha_4} 
 \right) \delta^{(32)}(\psi)\nonumber \\
\end{eqnarray}
Imposing eq. \eqref{totE} we can fix the components of $F^{(7|32)}$ in terms 
of those of $F^{(4)}$ as follows 
\begin{eqnarray}
\label{totH}
{\mathcal F}_{a_1 \dots a_7} &=& \epsilon_{a_1\dots a_7}^{~~~~~~ b_1 \dots b_4}  F^{(4)}_{b_1 \dots b_4}  \nonumber \\
{\mathcal F}_{a_1 \dots a_8}^\alpha &=& \epsilon_{a_1\dots a_8}^{~~~~~~ b_1 \dots b_3} C^{\alpha \beta} 
F^{(4)}_{b_1 \dots b_3 \beta}  \nonumber \\
\vdots~~~~~ 
&\vdots& ~~~~~\vdots \nonumber \\
{\mathcal F}_{a_1 \dots a_{11}}^{\alpha_1 \dots \alpha_4} &=& 
\epsilon_{a_1\dots a_{11}} C^{\alpha_1 \beta_1} \dots  C^{\alpha_4 \beta_4} 
F^{(4)}_{\beta_1 \dots \beta_4} 
\end{eqnarray}
where $C^{\alpha \beta}$ is the charge conjugation matrix. 

Eq. \eqref{totE} can be derived by an action as follows 
\begin{eqnarray}
\label{totI}
S = \int_{{\mathcal M}^{(11|32)}} \left( \frac12 F^{(7|32)}\wedge \star F^{(7|32)}  - 
F^{(4)} \wedge F^{(7|32)}  +{\mathcal L}^{(11|0)}(V, \psi, A^{(3)})\wedge \mathbb{Y}^{(0|32)}
\right) 
\end{eqnarray}
where ${\mathcal M}^{(11|32)}$ is the supermanifold on which we integrate the Lagrangian
${\mathcal L}^{(11|0)}(V, \psi, A^{(3)})$ (given in \cite{book} without the zero form 
$f_{a_1 \dots a_{4}}$) and $\mathbb{Y}^{(0|32)}$ is the PCO operator which allows us to 
convert the Lagrangian into an integral form. The PCO $\mathbb{Y}^{(0|32)}$ depends upon the 
vielbein and the gravitino field $\psi$, it is closed and not exact. Introducing the 3-form $A^{(3)}$ 
we can compute the equations of motion w.r.t. to $A^{(3)}$ and $ F^{(7|32)}$ as follows 
\begin{eqnarray}
\label{totL}
 F^{(7|32)} - \star F^{(4)} &=&0 \nonumber \\
 d F^{(7|32)} + \left( 
 \frac{\delta}{\delta A^{(3)}} {\mathcal L}^{(11|0)}(V, \psi, A^{(3)})\right) \wedge \mathbb{Y}^{(0|32)} &=& 0\,. 
\end{eqnarray}
where the derivative w.r.t. $A^{(3)}$ is the Euler-Lagrangian derivative. Since $\mathbb{Y}^{(0|32)}$ is independent of $A^{(3)}$ and, therefore, it can be extracted from the derivatives. In addition, since  
$\mathbb{Y}^{(0|32)}$ is closed, we can perform integration by parts in computing the functional derivatives of the action. The first equation implies the 
identifications of the d.o.f.'s of $F^{(4)}$ with those of $F^{(7|32)}$. Inserting this result into the first equation we 
get 
\begin{eqnarray}
\label{totM}
 d  \star F^{(4)} + 
 \left( \frac{\delta}{\delta A^{(3)}} {\mathcal L}^{(11|0)}(V, \psi, A^{(3)})\right) \wedge \mathbb{Y}^{(0|32)} &=& 0\,. 
\end{eqnarray}
Acting the inverse PCO $Z$ (which is also closed and not exact) we 
get 
\begin{eqnarray}
\label{totMA}
 d  \, Z \left( \star d A^{(3)} \right) + 
  \frac{\delta}{\delta A^{(3)}} {\mathcal L}^{(11|0)}(V, \psi, A^{(3)})  &=& 0\,. 
\end{eqnarray}
(at the moment assume that $Z$ acts only on $\mathbb{Y}^{(0|32)}$, but there should 
be more difficult cases). The expression is an $(8|0)$ form.  We notice that 
with the integral form $F^{(7|32)}$, the auxiliary zero form 
$f_{a_1 \dots a_{4}}$ is no longer needed. 

\section{Conclusions and Outlook}
In the present work, we have completed the analysis of the supergravity cocycles for 
different models extending the FDA beyond the superforms to integral forms and pseudoforms. 
We give general arguments of the structure of the FIDA, but we do not provide a complete 
mathematical analysis which might be very interesting along the lines of \cite{Fiorenza:2010mh}. 
We use the Hodge dual operator to complete some of the results obtained in previous work 
\cite{Cremonini:2022cdm}, and we also plan to work  the pseudo-form sector for those theories. 
A dual construction along \cite{FDAdual1} will be certainly interesting in order to develop the geometrical understanding of the underlying supergeometry and application to more general supermanifolds is a target for future 
works. 

\section*{Acknowledgements}

We would like to thank C.A. Cremonini, R. Catenacci, 
L. Castellani, R. D'auria, M. Trigiante, L. Ravera, R. Norris, L. Andrianopoli, for discussions and comments. 
The work is partially funded by the University of Eastern Piedmont with FAR-2019 projects.

\end{document}